\documentclass[]{pasj02} 
\usepackage{comment}
\usepackage{xcolor}
\usepackage{ulem}
\usepackage{textcomp}
\usepackage{multicol}
\usepackage{graphicx}
\usepackage{threeparttable}

\jyear{2026}
\Received{}%{yyyy/mm/dd}
\Accepted{}%{yyyy/mm/dd}
\begin{document} 

\title{AKARI near-infrared spectroscopy of circumstellar hydrocarbon dust associated with young stellar objects}

%%% begin:list of authors
% Do NOT capitalize all letters in "textsc".
\author{
 Yuki \textsc{Kuroda},\altaffilmark{1}\altemailmark\orcid{0000-0000-0000-0000} \email{y.kuroda@u.phys.nagoya-u.ac.jp} 
 Hidehiro \textsc{Kaneda},\altaffilmark{1}
 Natsuko \textsc{Izumi},\altaffilmark{1}
  Takuma \textsc{Kokusho},\altaffilmark{1}
 and 
 Shinki \textsc{Oyabu}\altaffilmark{2}
}
\altaffiltext{1}{Graduate School of Science, Nagoya University, Furo-cho, Chikusa-ku, Nagoya, Aichi 464-8602, Japan}
\altaffiltext{2}{Institute of Liberal Arts and Sciences, Tokushima University, 1-1 Minami-josanjima-cho,
Tokushima-shi, Tokushima, 770-8502, Japan}
%\altaffiltext{3}{C-Address of Institute}

%\footnotetext[$\dag$]{Present address: ....}

%%% end:list of authors

%% !!! Select 3 to 5 words from PASJ's key words !!! 
%% List of Key Words: https://academic.oup.com/pasj/pages/Pasj_Keywords 
%% "\KeyWords{ }" always has to be placed before ``\maketitle'' 

\maketitle

\begin{abstract}
We investigate the processing of the circumstellar polycyclic aromatic hydrocarbons (PAHs) associated with young stellar objects (YSOs) through their stellar activities. We analyze the AKARI near-infrared (2.6$-$4.4 \textmu m) long-slit spectral data of 10 YSOs, focusing on the aromatic feature at a wavelength of 3.3 \textmu m and the aliphatic features at 3.4$-$3.6 \textmu m. The spectral fitting results show large variations from target to target in the flux ratio of the aliphatic to aromatic features, the shape of the aliphatic feature profile, and the central wavelength of the aromatic feature. In particular, HD 97048 and Elias 1 show strong aliphatic features at wavelengths longer than 3.5 \textmu m, which is consistent with the hydrogenated nanodiamond features reported in previous studies. We find that the properties of the hydrocarbon dust vary under the influence of the YSO activity, where the degree of dust processing is likely to be related to the age of the YSO and the amount of the circumstellar dust. In addition, correlation is observed between the variations of the H$_2$O and CO$_2$ ice features and the aliphatic feature profiles, suggesting connection between the formation of CO$_2$ ice and the processing of aliphatic hydrocarbons. Finally, we find a significant correlation between the variations in the central wavelength of the aromatic feature and the shape of the aliphatic feature profile. These findings may be consistent with a tentative scenario in which circumstellar PAHs are converted to nanodiamonds through irradiation by high-energy particles produced by YSO activity.

\end{abstract}

%\pagewiselinenumbers 

\section{Introduction}

 \begin{table*}
  \tbl{Characteristics of the sample YSOs in the present study.}{
  \begin{tabular}{lccccl}
      \hline
      Object name & Distance$^*$ (pc) & Spectral type$^*$ & Category & Age$^*$ (Myr) \\ 
      \hline
      Elias 1 & 118 $\pm$ 2$^{(1)}$ & A0$^{(3)}$ & Herbig Ae/Be & 5$^{(13)}$  \\
      HD97048 & 185 $\pm$ 1$^{(1)}$ & A0$^{(4)}$ & Herbig Ae/Be & 4.7$-$6.1$^{(14)}$ \\
      BD+40 4124 & 909 $\pm$ 26$^{(1)}$ & B2$^{(5)}$ & Herbig Ae/Be & $<$0.5$^{(15)}$  \\
      TY CrA$^\dag$& 137 $\pm$ 3$^{(1)}$ & B9$^{(6)}$ & Herbig Ae/Be & 3$-$5$^{(16)}$ &  \\
      CD-42 11721 & 1667 $\pm$ 152$^{(1)}$ & B0$^{(7)}$ & Herbig Ae/Be & 0.023$^{(17)}$ &  \\
      SU Aur & 159 $\pm$ 1$^{(1)}$ & G2$^{(8)}$ & T Tauri & 2.2$^{(18)}$  \\
      IRAS 16226-2420 & 217 $\pm$ 19$^{(1)}$ & M2$^{(9)}$ & T Tauri & 3.8$^{(19)}$ \\
      CPD-36 6759 & 135 $\pm$ 1$^{(1)}$ & F8$^{(10)}$ & Herbig Ae/Be & 8$^{(20)}$ \\
      LZK 12 & 303 $\pm$ 3$^{(2)}$ & B5$^{(11)}$ & YSO$^\ddag$ & 0.6$^{(21)}$ \\
      2MASS J18034104-2422413 & 909 $\pm$ 114$^{(1)}$  & K0-2$^{(12)}$ & T Tauri & $<$1$^{(12)}$  \\
      \hline 
      \end{tabular}}
      \label{tab:target}
\begin{tabnote}
\footnotemark[$*$] Sources: (1) Gaia collaboration (2018). (2) Cantat$-$Gaudin $\&$ Anders (2020). (3) Skiff (2014). (4) Irvine $\&$ Houk (1977). (5) Mora et al. (2001). (6) Vieira et all. (2003). (7) Lopes et al. (1992).
(8) Johns et al. (1995). 
(9) Wilking et al. (2005). (10) Coulson $\&$ Walther (1995). (11) Straizys et al. (2002). (12) Arias et al. (2007).
(13) Hamidouche (2010). (14) van der Marel et al. (2019) (15) Strom et al. (1972) (16) Va{\v{n}}ko et al. (2013). (17) Vioque et al. (2018). (18) Jones et al. (2012). (19) Erickson et al. (2011). (20) Bae (2017).  (21) Flores et al. (2024). 
\\
\footnotemark[$\dag$] TY CrA is a multiple system, and several previous studies suggest that its primary star may have already reached the main sequence (e.g., Casey et al. 1998; Va{\v{n}}ko et al. 2013)\\ 
\footnotemark[$\ddag$] No information available on the category.\\
\end{tabnote}
\end{table*}

\begin{table*}[htbp]
\caption{Initial values and the range of the parameters for the dust features in the spectral model fitting.}
\label{list_parameters}
\centering
\begin{tabular}{ccc}
\hline
Feature & Central wavelength (\textmu m) & Feature width (FWHM, \textmu m) \\
 \hline
Aromatic & 3.3 (3.28$-$3.32) & 0.046 (0.040$-$0.050) \\
Aliphatic 1 & 3.41 (3.39$-$3.44) & 0.03 (0.025$-$0.1) \\
Aliphatic 2 & 3.46 (3.44$-$3.49) & 0.03 (0.025$-$0.035) \\
Aliphatic 3 & 3.51 (3.49$-$3.53) & 0.03 (0.025$-$0.035) \\
Aliphatic 4 & 3.56 (3.53$-$3.59) & 0.03 (0.025$-$0.035) \\
\hline
$\mathrm{H_2O}$ ice & 3.05 (2.90$-$3.20) & 0.13 (0.05$-$) \\
$\mathrm{CO_2}$ ice & 4.27 (4.22$-$4.32) & 0.15 (0$-$) \\
\hline    
\end{tabular}
\end{table*}

A young stellar object (YSO) refers to an early stage in stellar evolution during the process of star formation, and studies of YSOs constitute an important field in infrared (IR) astronomy. As YSOs evolve, it is known that their high-energy activities increase, including flare events driven by magnetic reconnection at the stellar surface, as well as magnetic interactions between the star and its circumstellar disk (e.g., \cite{and94}; \cite{fei99}). High-energy radiation from evolved YSOs can alter the physical conditions of the circumstellar disk and affect the properties of the surrounding circumstellar materials. These effects may influence the chemical evolution during subsequent planet formation, and potentially even the emergence of life on planets.

IR spectra reveal the presence of a wide variety of materials surrounding YSOs. Among them, in the wavelength range from the near-IR at $\sim$3 \textmu m to the mid-IR at $\sim$17 \textmu m, a number of strong dust emission features arising from polycyclic aromatic hydrocarbons (PAHs) are observed (e.g., \cite{dra07}; \cite{tie08}), although it is known that a relatively small fraction of low-mass YSOs are detected in the PAH emission (e.g., \cite{gee09}; \cite{goto09}). In the near-IR, a prominent PAH feature appears at 3.3 \textmu m, while band features due to aliphatic hydrocarbons are found on the longer-wavelength side at 3.4$-$3.6 \textmu m. All these hydrocarbon dust features in the near-IR are attributed to C-H vibrations. % In addition, multiple hydrogen recombination lines are detected, including Br$\alpha$ at 4.05 \textmu m. Broad absorption due to H$_2$O ice centered around 3 \textmu m, as well as relatively narrow absorption from CO$_2$ ice at 4.3 \textmu m, are also observed. 

Processing of hydrocarbon dust can be studied through near-IR spectroscopy. In general, aliphatic C-H bonds are more fragile than aromatic C-H bonds, and therefore, in harsh environments with strong high-energy radiation such as ultraviolet (UV) and X-rays, the ratio of the aliphatic to aromatic feature intensity is expected to decrease (e.g., \cite{yan13}; \cite{kon26}). In contrast, observations of nearby galaxies with the AKARI IR satellite have shown that aliphatic features become relatively strong in more extreme environments (e.g., \cite{yama12}), suggesting that hydrocarbon dust is being fragmented by processes such as interstellar shocks (\cite{jones13}). Hydrogenated amorphous carbon, which is a representative form of hydrocarbon dust, contains a variety of C-H bonding configurations, and thus the central wavelengths of the corresponding band features vary depending on the bonding environment of the carbon atom associated with the hydrogen atom (e.g., \cite{kwok07}). Recent spectroscopic observations of YSOs with the James Webb Space Telescope (JWST) show detailed properties of PAHs in protoplanetary disks (e.g., \cite{ban25}, \cite{stu24}). JWST reveals the processing of circumstellar PAHs with the disk evolution associated with the central UV radiation (e.g., \cite{aru25}). Besides astronomical observations, IR analyses of the samples returned from Ryugu by Hayabusa2 also reveal the presence of the aromatic and the aliphatic spectral features, and that even within a single grain, absorption features can vary depending on the location within the sample analyzed, indicating that the material has undergone processing likely due to irradiation by the Sun (\cite{pilo21}).

An extreme example of hydrocarbon dust processing is the production of nanodiamonds. Previous studies have identified 4 extrasolar objects that exhibit clear nanodiamond emission features: 3 Herbig Ae/Be stars, Elias 1 (\cite{whit84}), MWC297 (\cite{tera01}), and HD97048 (\cite{whit83}) as well as 1 post-AGB star, HR4049 (\cite{geb89}). Both Elias 1 and HD97048 show the typical PAH emission feature at 3.3 \textmu m, as well as hydrogen recombination lines such as Br$\alpha$ and Br$\beta$. The most prominent feature is the aliphatic emission features at 3.4$-$3.6 \textmu m, with a particularly strong feature centered at 3.53 \textmu m. These features are interpreted as emission from hydrogenated nanodiamonds (i.e., diamond surfaces terminated with hydrogen), formed through the processing of hydrocarbon dust by high-energy phenomena associated with the stellar activity (\cite{vank02}; \cite{goto09}). Nanodiamonds are also found within the Solar System, in presolar dust grains contained in meteorites (e.g., \cite{kwok07}). 
 
 The aim of the present study is to investigate how hydrocarbon dust in the circumstellar environments of YSOs is processed as functions of various tracers of the YSO activity, using near-IR 2.6$-$4.4  \textmu m spectroscopic observational  data obtained with AKARI. By analyzing the long-slit spectra of 10 YSOs, we systematically examine variations in the spectral features. AKARI near-IR wavelength range covers not only the hydrocarbon dust features, but also the hydrogen recombination lines including Br$\alpha$ at 4.05 \textmu m and the H$_2$O and CO$_2$ ice absorption features centered around 3 \textmu m and 4.3 \textmu m, respectively, which enable diagnostics of the circumstellar environment and allow us to explore their relationship with the dust processing. 
  
%\noindent IMPORTANT NOTICE\\
%1. Manuscript for submission must be in the same format as a published papers. \\
%2. Line numbers should be added to the manuscript. \\
%3. Do NOT use ``\verb|\def|, \verb|\renewcommand|''.\\
%4. Do NOT redefine commands provided by pasj02.cls.  

\section{Sample selection and data analysis}\label{sec:2}
%%用いたデータとデータ解析
\begin{figure*}[tbp]
	\centering
	\begin{minipage}{0.45\linewidth}
		\centering
		\includegraphics[width=0.8\linewidth]{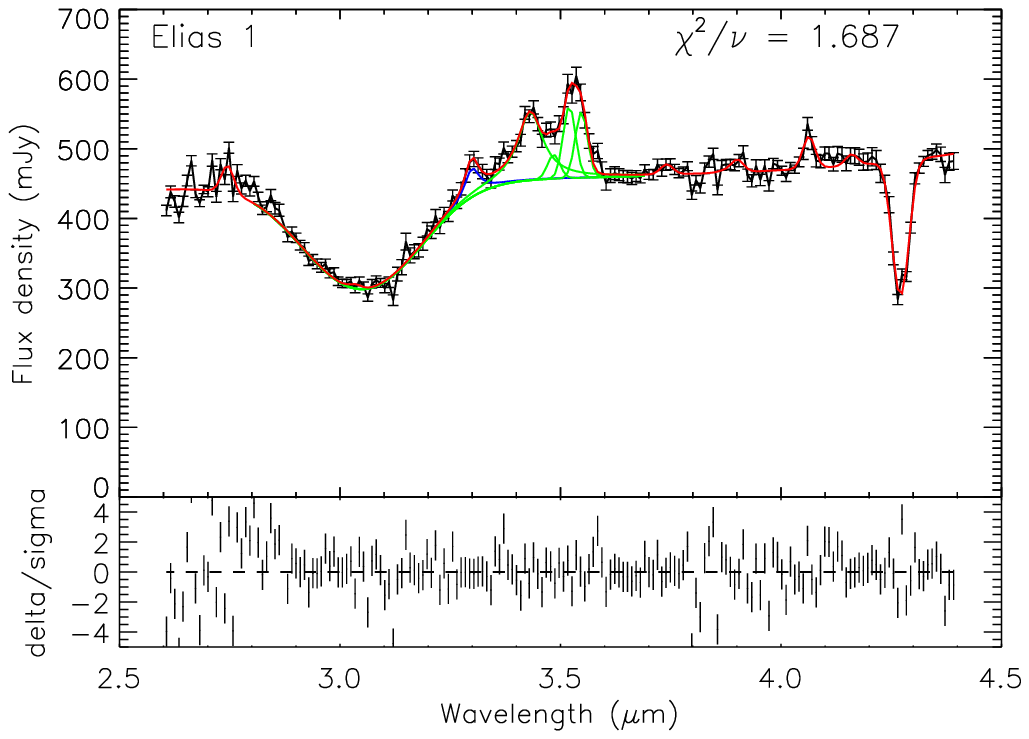}
		\includegraphics[width=0.8\linewidth]{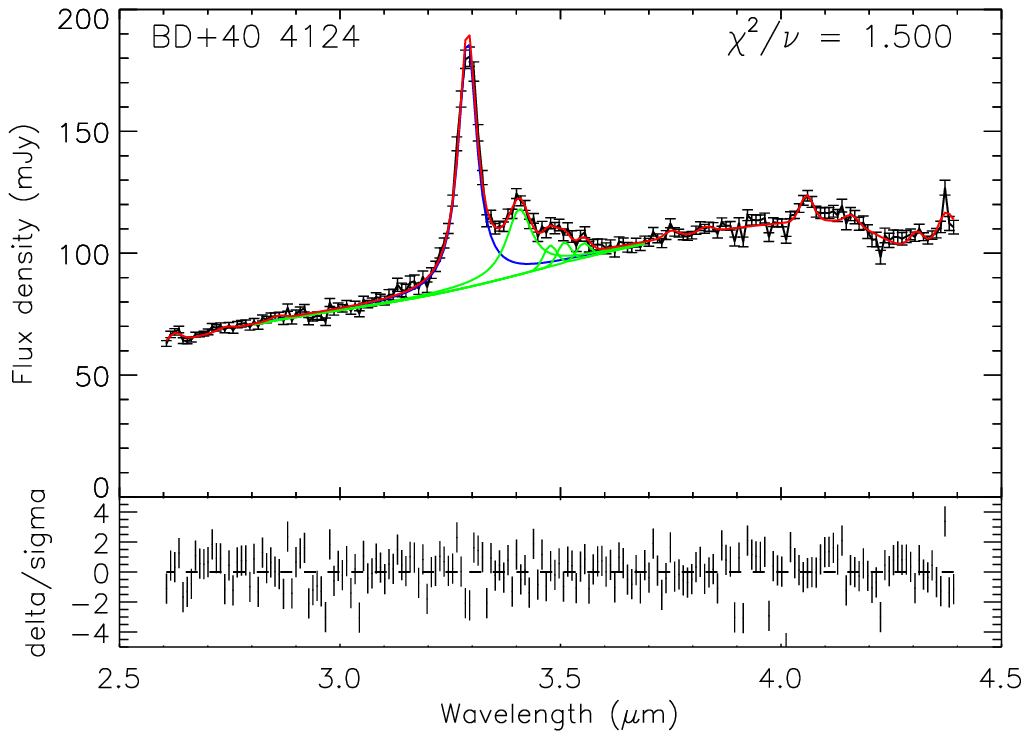}
		\includegraphics[width=0.8\linewidth]{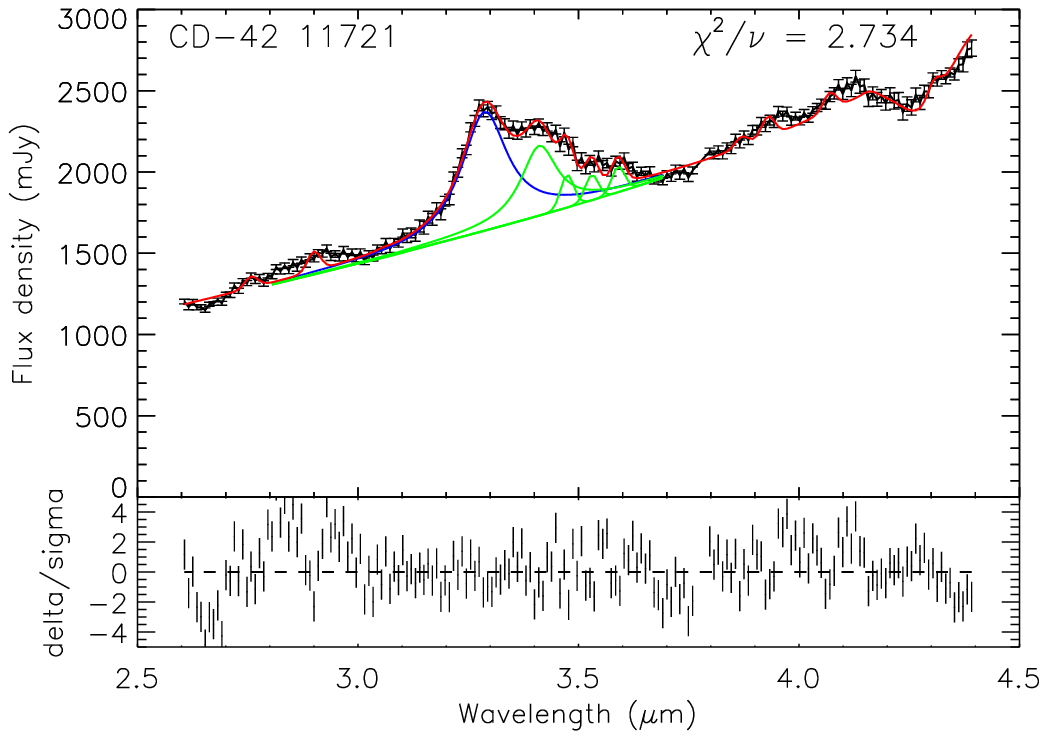}
		\includegraphics[width=0.8\linewidth]{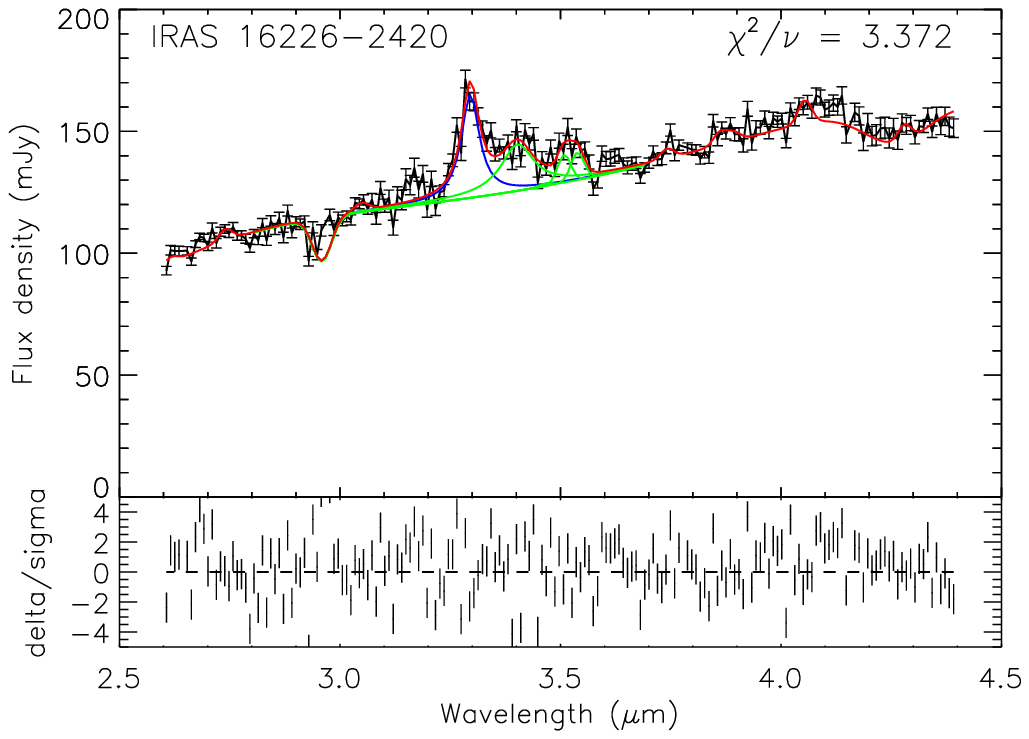}
		\includegraphics[width=0.8\linewidth]{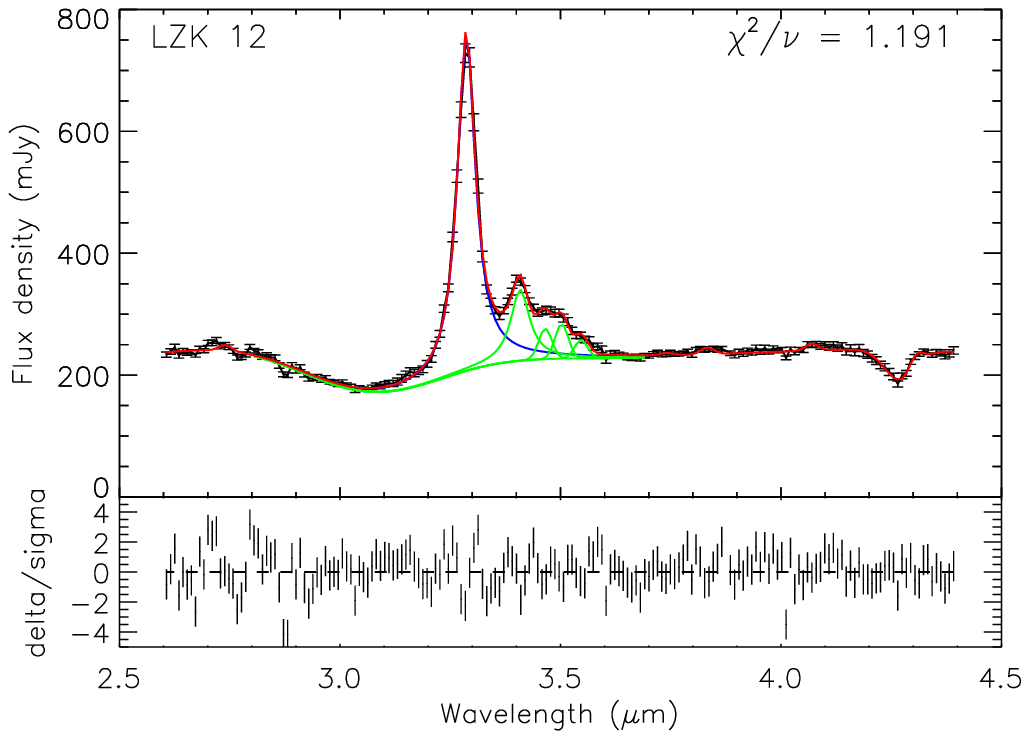}
	\end{minipage}
	\begin{minipage}{.45\linewidth}
		\centering
		\includegraphics[width=0.8\linewidth]{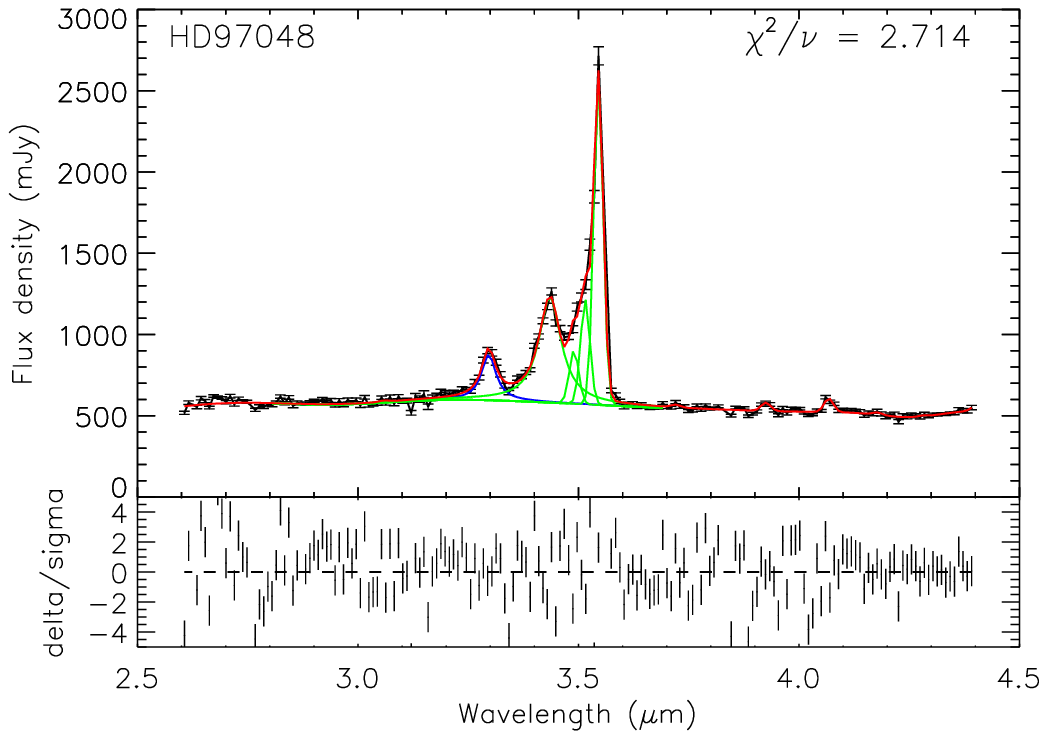}
		\includegraphics[width=0.8\linewidth]{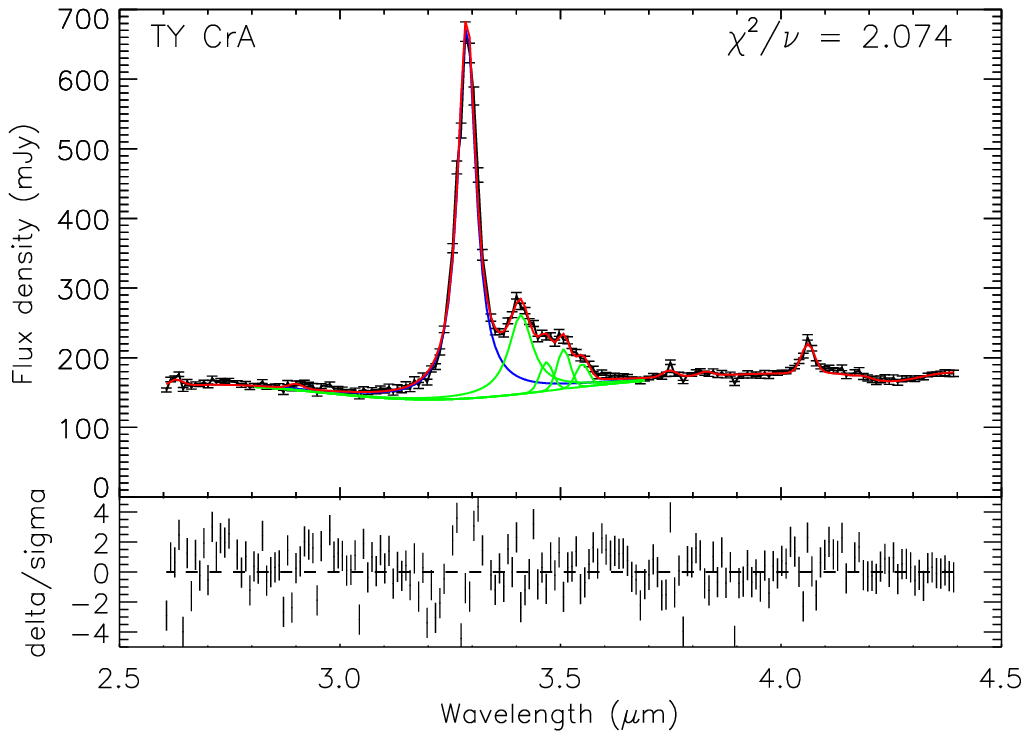}
		\includegraphics[width=0.8\linewidth]{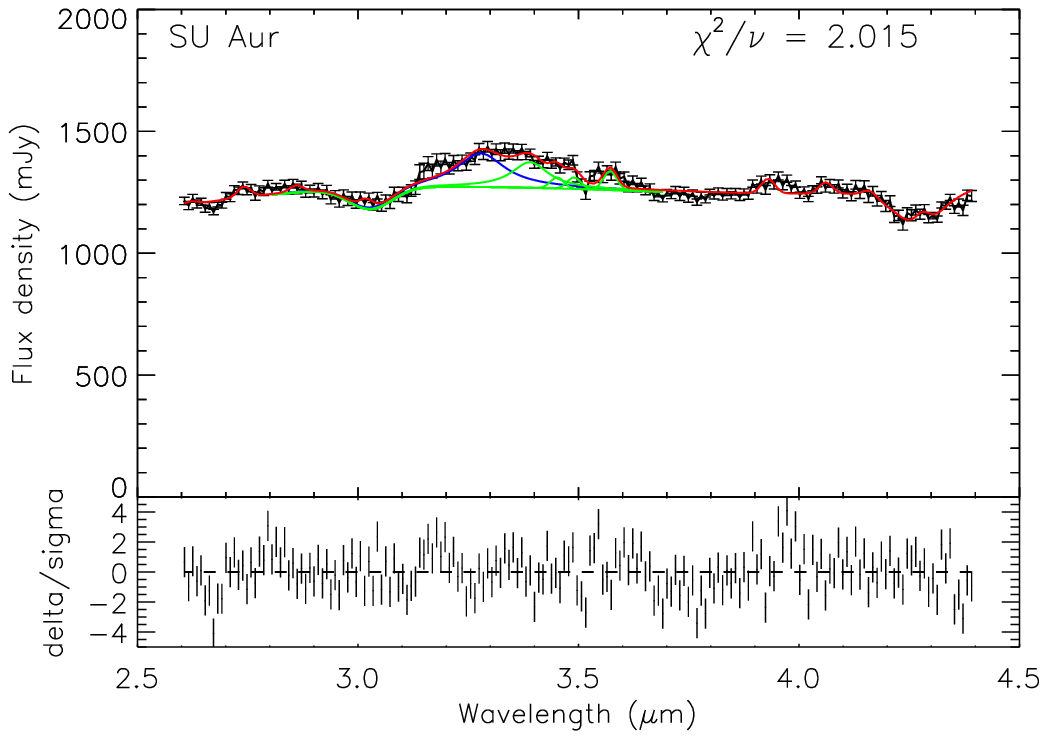}
		\includegraphics[width=0.8\linewidth]{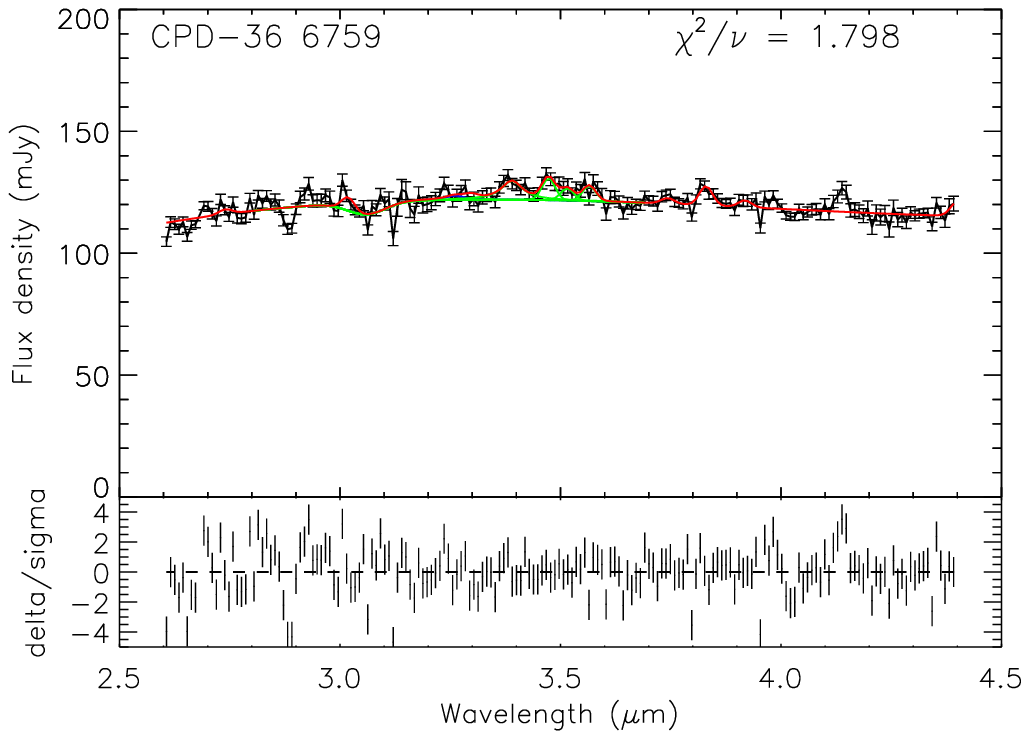}
		\includegraphics[width=0.8\linewidth]{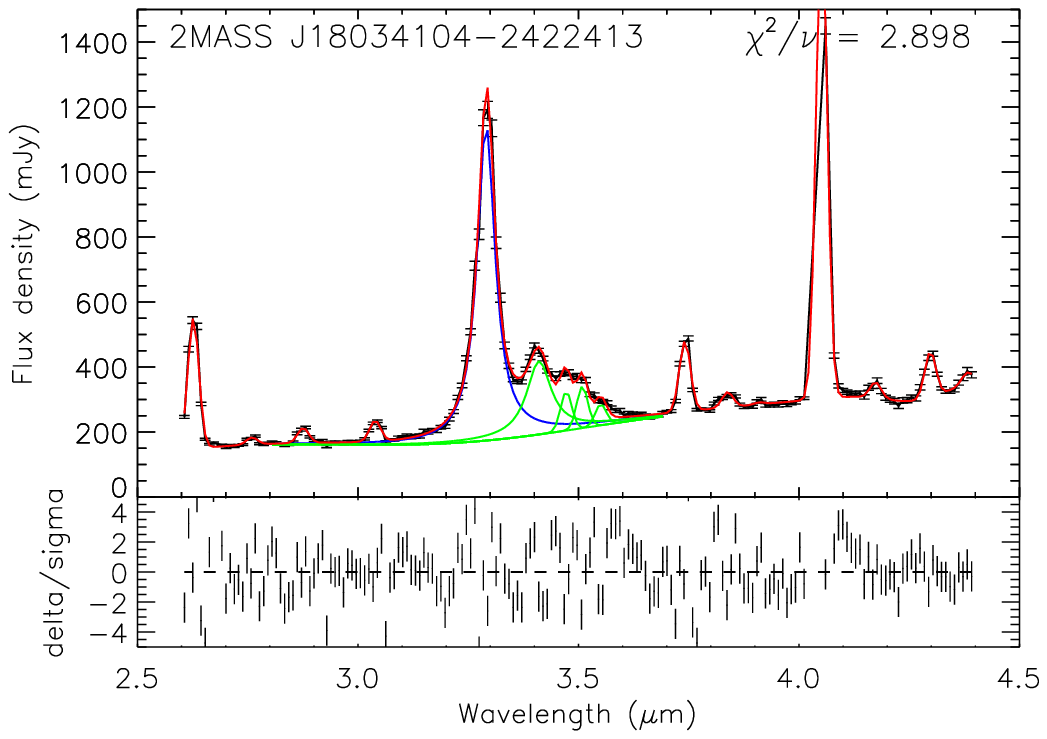}
	\end{minipage}
	
	\vspace{3mm}
	
	\caption{Results of the model fitting to the spectra for the inner 10\farcs5 regions of the sample YSOs. The black and red lines represent the observed spectrum and the best-fit model, respectively. The blue and green lines show the aromatic and aliphatic components, respectively. The fitting residuals normalized by the errors are shown in the bottom panels. {Alt text: Ten panels showing the results of the spectral fitting. }} 
	\label{figure1}
\end{figure*}

		\begin{table*}[htb]
	\centering
        \caption{Fitting results for the fluxes of the aromatic and aliphatic emission features in the inner and outer regions.}
	\label{tab:aliaro}
		\begin{tabular}{lllll}
		\hline
		Object name &\multicolumn{2}{c}{Inner region} &\multicolumn{2}{c}{Outer region$^*$} \\
			& Aromatic & Aliphatic & Aromatic & Aliphatic \\
			& \multicolumn{4}{c}{($\times 10^{-15}$ W m$^{-2}$)}  \\
		\hline\hline
		Elias 1 & 0.68 $\pm$ 0.12 & 4.89 $\pm$ 0.18 &0.026 $\pm$ 0.005 & 0.145 $\pm$ 0.006 \\
		HD97048 & 5.63 $\pm$ 0.21 & 38.27 $\pm$ 0.59 & 0.216 $\pm$ 0.007 & 1.262 $\pm$ 0.013\\
		BD+40 4124 & 2.384 $\pm$ 0.074 & 1.030 $\pm$ 0.043 & 0.906 $\pm$ 0.010 & 0.400 $\pm$ 0.008 \\
		TY CrA & 13.88 $\pm$ 0.30 & 4.564 $\pm$ 0.092 & 7.33 $\pm$ 0.12 & 3.34 $\pm$ 0.03\\
		CD-42 11721 & 39.14 $\pm$ 0.77& 21.14 $\pm$ 0.61 & 12.97 $\pm$ 0.11 & 5.58 $\pm$ 0.08\\
		SU Aur & 8.24 $\pm$ 0.56 & 6.28 $\pm$ 0.38 & 0.008 $\pm$ 0.006 & 0.133 $\pm$ 0.005 \\
		IRAS 16226-2420 & 0.935 $\pm$ 0.045 & 0.976 $\pm$ 0.049& 1.060 $\pm$ 0.012 & 0.617 $\pm$ 0.014 \\
		CPD-36 6759 & 0.023 $\pm$ 0.029 & 0.371 $\pm$ 0.039 & 0.027 $\pm$ 0.004 & 0.052 $\pm$ 0.007\\
		LZK 12 & 13.14 $\pm$ 0.16 & 3.628 $\pm$ 0.094 & 8.52 $\pm$ 0.07 & 2.597 $\pm$ 0.020 \\
		2MASS J18034104-2422413 & 27.58 $\pm$ 0.74 & 10.50 $\pm$ 0.19 & 17.28 $\pm$ 0.16 & 5.31 $\pm$ 0.06 \\
	\hline
		\end{tabular}
		\begin{tabnote}
\footnotemark[$*$] Fluxes multiplied by a factor of 7/23 considering the difference in the slit aperture size between the inner and outer regions.
\end{tabnote}
\end{table*}

\begin{table*}[htb]
	\centering
        \caption{Line fluxes of the hydrogen recombination line Br$\alpha$, and the EWs of the H$_2$O and CO$_2$ ice absorption features in the inner regions.}
	\label{tab:lineice}
		\begin{tabular}{llllll}
		\hline
		Object name & Br$\alpha$ line flux  & H$_2$O ice EW & CO$_2$ ice EW& \multicolumn{2}{c}{Continuum flux} \\
					&	&	&	& 2.6$-$2.8 $\mu$m & 3.75$-$3.95 $\mu$m \\
		                    &\multicolumn{1}{c}{($\times 10^{-15}$ W m$^{-2}$)}& \multicolumn{1}{c}{(nm)} & \multicolumn{1}{c}{(nm)}& \multicolumn{2}{c}{($\times 10^{-15}$ W m$^{-2}$)} \\
		\hline\hline
		Elias 1 & 0.303 $\pm$ 0.041 & 108.4 $\pm$ 1.9 & 19.14 $\pm$ 0.60 & 36.3 $\pm$ 0.1 & 18.51 $\pm$ 0.07  \\
		HD97048 & 0.436 $\pm$ 0.039 & 25.1 $\pm$ 4.4 & 11.1 $\pm$ 2.3 & 46.8 $\pm$ 0.2 & 21.67 $\pm$ 0.08  \\
		BD+40 4124 & 0.0671 $\pm$ 0.0099* & 53 $\pm$ 11 & (23.9 $\pm$ 3.6)$^\dag$ & 5.51 $\pm$ 0.02 & 4.33 $\pm$ 0.02  \\
		TY CrA & 0.295 $\pm$ 0.016 & 107.7 $\pm$ 8.5 & 14.3 $\pm$ 2.3 & 13.22 $\pm$ 0.05 & 6.95 $\pm$ 0.03 \\
		CD-42 11721 & $-$0.37 $\pm$ 0.16* & (2.3 $\pm$ 1.3)$^\dag$ & (15.7 $\pm$ 1.5)$^\dag$ & 102.7 $\pm$ 0.3 & 85.2 $\pm$ 0.2 \\
		SU Aur & 0.259 $\pm$ 0.073 & 10.2 $\pm$ 1.2 & 10.4 $\pm$ 1.0 & 100.4 $\pm$ 0.3 & 49.7 $\pm$ 0.1 \\
		IRAS 16226-2420 & 0.043 $\pm$ 0.014* & (8.97 $\pm$ 0.62)$^\dag$ & (18.0 $\pm$ 2.3)$^\dag$ & 8.42 $\pm$ 0.03 & 5.75 $\pm$ 0.02 \\
		CPD-36 6759 & 0.00 & (2.91 $\pm$ 0.58)$^\dag$ & 0.00 & 9.41 $\pm$ 0.03 & 4.80 $\pm$ 0.02 \\
		LZK 12 & 0.084 $\pm$ 0.021 & 106.3 $\pm$ 3.3 & 19.4 $\pm$ 1.2 & 19.66 $\pm$ 0.07 & 9.36 $\pm$ 0.03 \\
		2MASS J18034104-2422413 & 8.513 $\pm$ 0.082 & 180 $\pm$ 13 & 21.1 $\pm$ 3.0 & 18.37 $\pm$ 0.08 & 11.64 $\pm$ 0.04 \\
	\hline
		\end{tabular}
		\begin{tabnote}
\footnotemark[$*$] The fluxes are derived using fits over a local wavelength range around the line.
\footnotemark[$\dag$] The values may result from fitting spurious spectral structures and are therefore excluded from the result and discussion.
\end{tabnote}

\end{table*}

Table 1 lists the sample YSOs analyzed in the present study, along with their distances, spectral types, and ages. These objects correspond to the YSOs which were observed with the AKARI near-IR long-slit ($\sim$1 arcmin in length) spectroscopy mode, showing the PAH emission or, at least, a hint of excess emission above the continuum at wavelengths of 3.3$-$3.6 \textmu m. All the observations of the sample YSOs were carried out during Phase 3 of AKARI after the depletion of liquid helium. We processed the observational data using the official IRC data reduction  pipeline for Phase 3 (IRC Spectroscopy Toolkit Version 201821203)\footnote{https://www.ir.isas.jaxa.jp/AKARI/Observation/support/IRC/}. In order to treat spatially-extended emissions associated with the YSOs, we did not adopt the aperture corrections that were specifically derived for point-like sources and automatically applied in the data reduction pipeline. In Phase 3, the detector temperature was relatively high ($\sim$50 K), and the spectral data contained hot pixels; therefore, we identified, removed, and interpolated the hot pixels. For CD-42 11721 and SU Aur, the spectral intensities in their inner regions were so strong as to cause signal saturation, and thus, instead of using the standard long-exposure data (44.4 s), we adopted the short-exposure data (4.68 s).

We apply model fitting to the spectral data obtained through the above procedures for the wavelength range of 2.6 to 4.4 \textmu m.  Among the prominent features observed in the near-IR spectra, the primary focus of the present study is on the PAH emission features which arise from aromatic hydrocarbons at 3.3 \textmu m and from aliphatic hydrocarbons at 3.4$-$3.6 \textmu m. In order to characterize those emission features, we adopt a Drude profile for the 3.3 \textmu m aromatic feature ($F_{\nu, \rm{aro}}$) and a Lorentzian and 3 Gaussian profiles for the 3.4$-$3.6 \textmu m aliphatic feature ($F_{\nu, \rm{ali}}$; e.g., \cite{slo97}; \cite{kwok11}), following the procedure adopted in \citet{kon26}. Since we expect the overall shape of the aromatic and the aliphatic feature profile to significantly vary depending on the chemical structure of the hydrocarbons, we treat the central wavelengths and the widths of all the components as free parameters within the fitting ranges listed in table \ref{list_parameters}. We define the fitting wavelength ranges so as not to overlap between the 4 aliphatic components, allowing for a consistent treatment of both typical linear-chain aliphatic features and diamond-like aliphatic features.

For the gas emission lines, we consider 12 lines (H\emissiontype{I} Br$\mathrm{\beta}$, H\emissiontype{I} Pf12, H\emissiontype{I} Pf$\mathrm{\eta}$, H\emissiontype{I} Pf$\mathrm{\epsilon}$, H\emissiontype{I} Pf$\mathrm{\delta}$, H\emissiontype{I} Pf$\mathrm{\gamma}$, H$\mathrm{_2}$ S(13), H\emissiontype{I} Hu15, H\emissiontype{I} Br$\mathrm{\alpha}$, H\emissiontype{I} Hu13, He\emissiontype{I}, and H\emissiontype{I} Hu12), all of which are treated as narrow Gaussian lines ($F_{\nu, {\rm gas}}$). Since one of the hydrogen recombination lines, Pf$\delta$, overlaps with the 3.3 \textmu m aromatic feature, it is difficult to estimate the flux of the aromatic feature with the spectral resolution of AKARI/IRC, and therefore we fix its amplitude at 0.69 times that of Pf$\gamma$ in the spectral fitting, based on the Case B recombination (\cite{hum87}). As for the ice absorption features, H$_2$O and CO$_2$ ices are considered and modeled using Gaussian functions ($\tau_\nu$). We fit the continuum using either a combination of power-law and Planck functions or a fourth-order polynomial function, and adopt the function providing the better fit ($F_{\nu, \rm{cont}}$). Specifically, the former is used for Elias 1, CD-42 11721, CPD-36 6759, and SU Aur, while the latter for the other 6 targets. Hence the flux densities of the observed spectra are expressed as, 
	\begin{equation}
	F_{\nu} = (F_{\nu, {\rm aro}} + F_{\nu, {\rm ali}} + F_{\nu, {\rm gas}} + F_{\nu, {\rm cont}} )\times\exp(-\tau_{\nu}).
	\end{equation}
Using the above model, we carry out the spectral fitting in the following three steps: first, in order to determine the continuum shape, we perform the fitting using the above continuum models for the following wavelength ranges which are relatively free from the dust emission features, the gas emission lines, and the ice absorption features: 2.5$-$2.6 \textmu m, 2.7$-$2.75 \textmu m, 3.6$-$3.7 \textmu m, 3.75$-$3.95 \textmu m, 4.1$-$4.15 \textmu m, and 4.35$-$4.4 \textmu m. As a next step, we repeat the spectral fitting with the model which includes the dust emission features, gas emission lines, and ice absorption features using the above functions, while keeping the continuum parameters fixed at the values obtained in the previous step. Finally, we perform the third fitting in which the continuum parameters, as well as the amplitudes of the dust emission features, gas emission lines, and ice absorption features, are allowed to vary freely.

\begin{figure*}
 \centering
 \begin{minipage}[b]{0.49\linewidth}
  \centering
  \includegraphics[width=10cm]{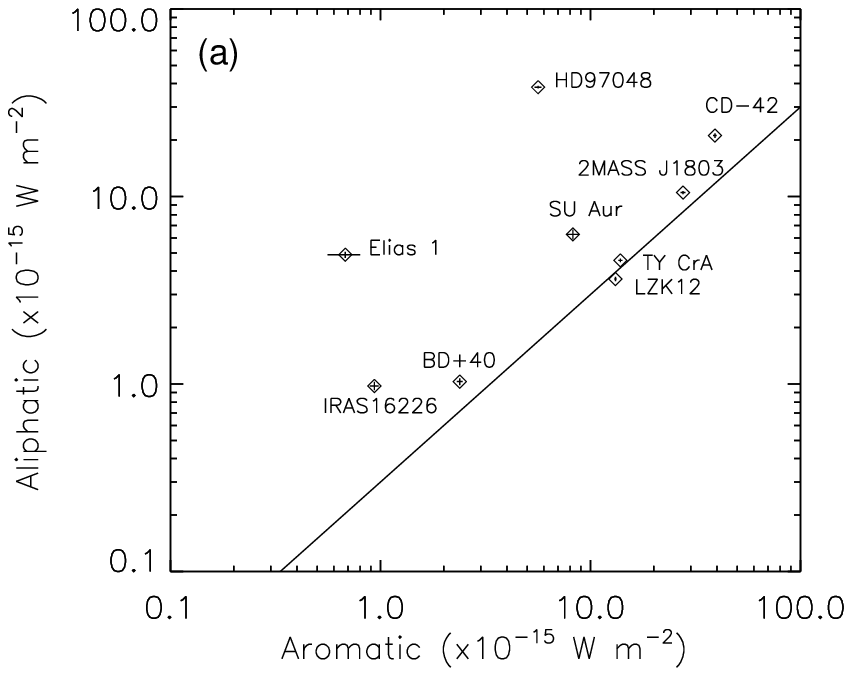} 
\end{minipage}
\begin{minipage}[b]{0.49\linewidth}
 \centering
  \includegraphics[width=10cm]{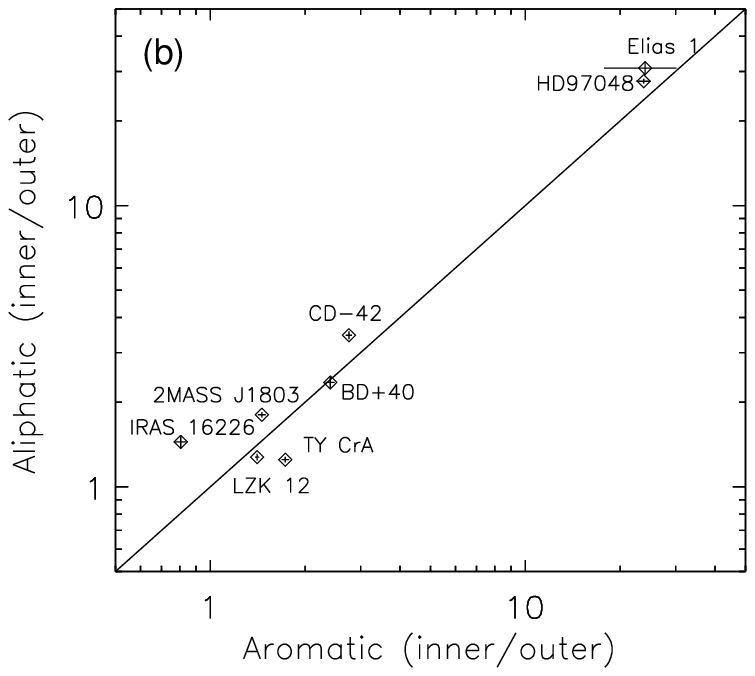} 
\end{minipage}
\caption{(a) Relationship between the aliphatic and aromatic feature fluxes for the inner 10\farcs5 region of each YSO. The solid line represents the relation typical of the interstellar hydrocarbon dust in a star-forming galaxy ($y=0.3x$; \cite{kon26}). (b) Relationship between the aliphatic and the aromatic flux ratios between the inner (<10\farcs5) and outer (10\farcs5$-$45\farcs0) regions. The solid line indicates $y=x$, where the flux ratio of the aliphatic to aromatic features remain unchanged from the outer to the inner region. {Alt text: Two scatter plots.}}
\label{figure2}
\end{figure*}

\section{Results}\label{sec:3}
%%結果
\subsection{Overall model fitting to the total near-infrared spectra}\label{ssec:31}

Figure \ref{figure1} shows the results of the model fitting to the total near-IR spectra of the 10 sample YSOs for their central regions, using the spectral model described in section 2. For each object, spectra are obtained not only for the inner region ($\le$ 10\farcs5, including the central source) but also for the outer region (10\farcs5$-$45\farcs0). The quality of the fits varies among the targets, particularly in the continuum at wavelengths longer than 4 \textmu m. For the 3.2$-$3.7 \textmu m range where the hydrocarbon features are present, however, the observed spectra of all the targets are reproduced fairly well by the model. The fluxes of the aromatic and aliphatic emission features derived from the inner and outer regions are summarized in table \ref{tab:aliaro}, where the fluxes of the outer regions are multiplied by a factor of 7/23, considering the difference in the slit aperture size between the inner and outer regions. The table shows that both aromatic and aliphatic hydrocarbon emission features are significantly (S/N $\ge$ 5) detected from the inner region of every target, except for CPD-36 6759 in the aromatic emission. Hence all the targets except CPD-36 6759 are included in the relevant scatter plots shown below. 
\begin{figure*}[tbp]
	\centering
	\begin{minipage}{0.45\linewidth}
		\centering
		\includegraphics[width=0.8\linewidth]{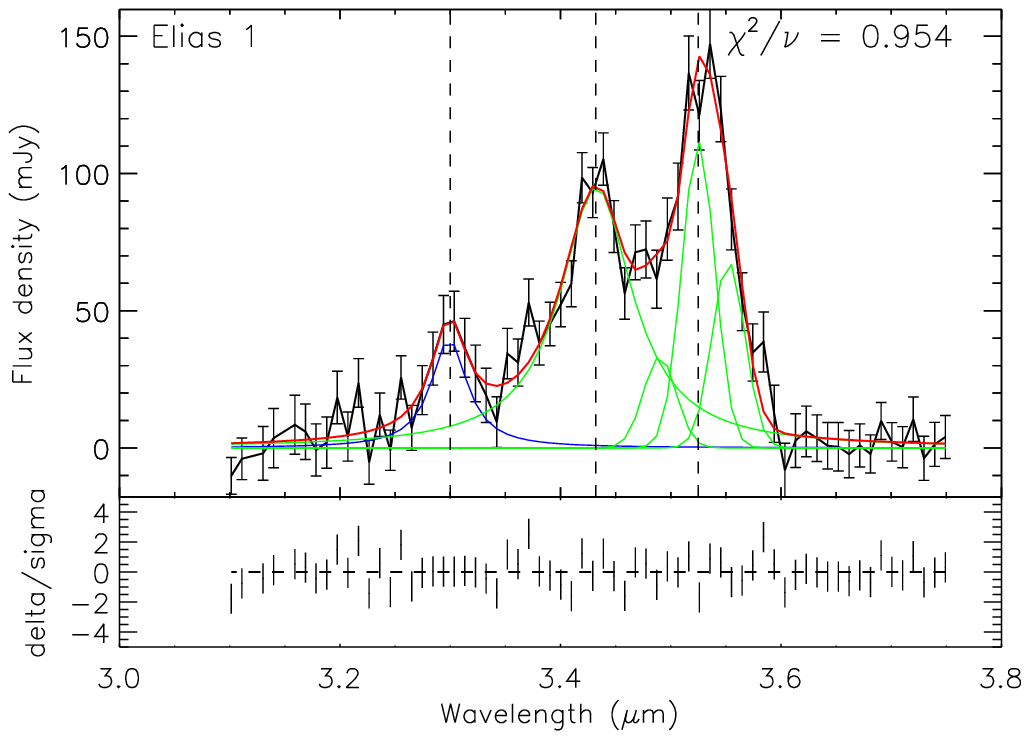}
		\includegraphics[width=0.8\linewidth]{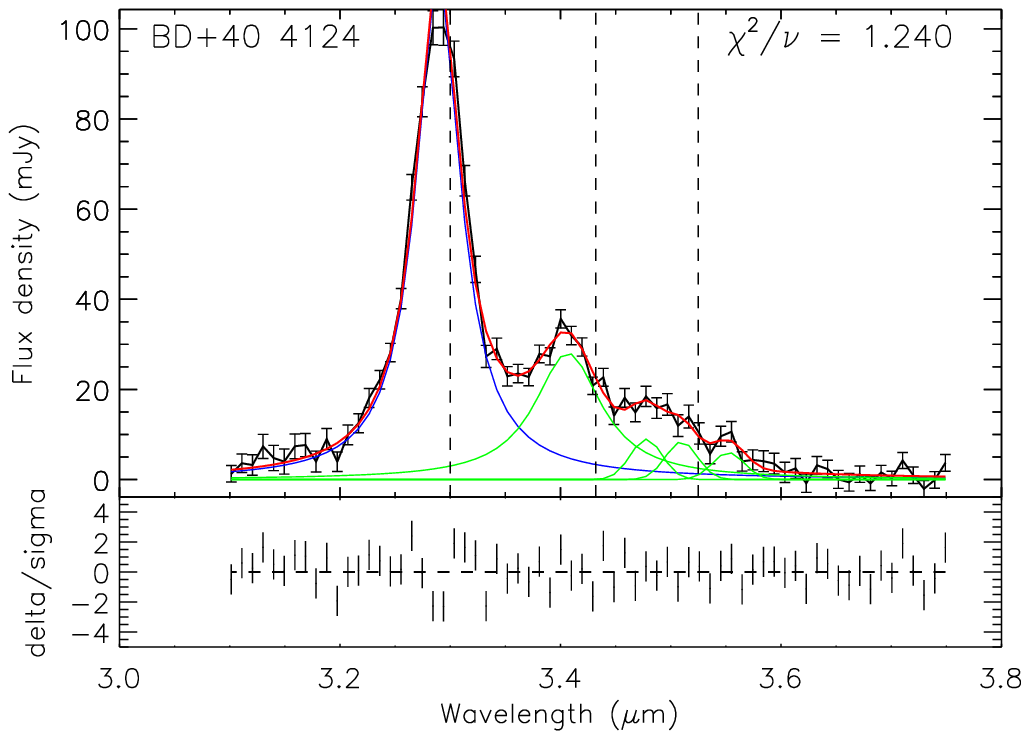}
		\includegraphics[width=0.8\linewidth]{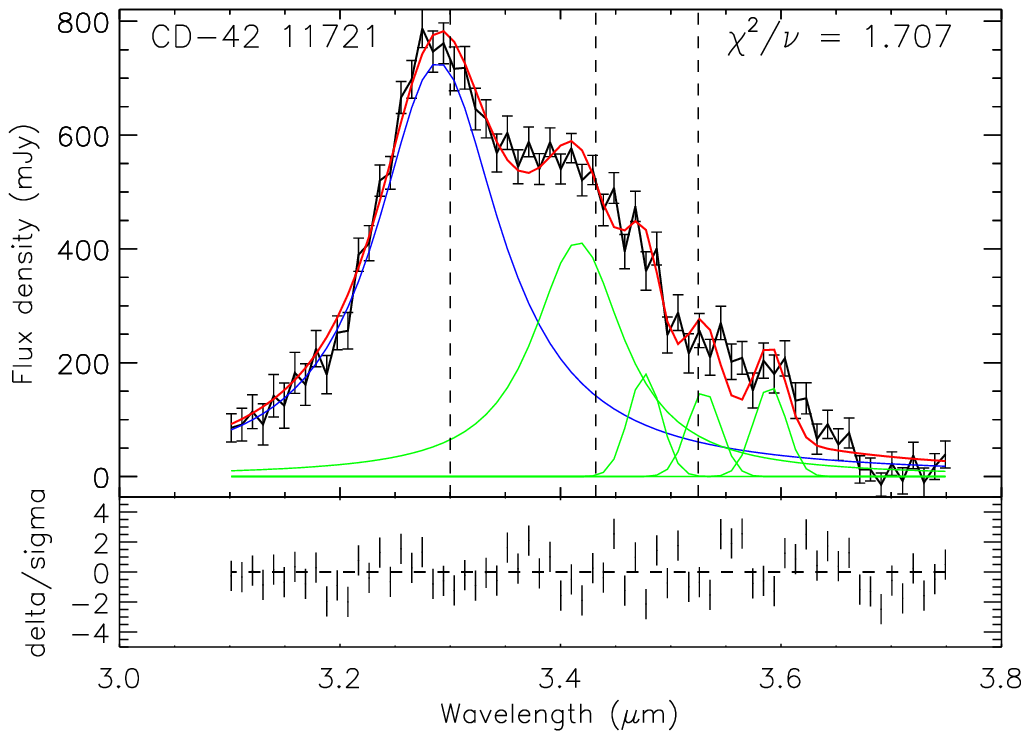}
		\includegraphics[width=0.8\linewidth]{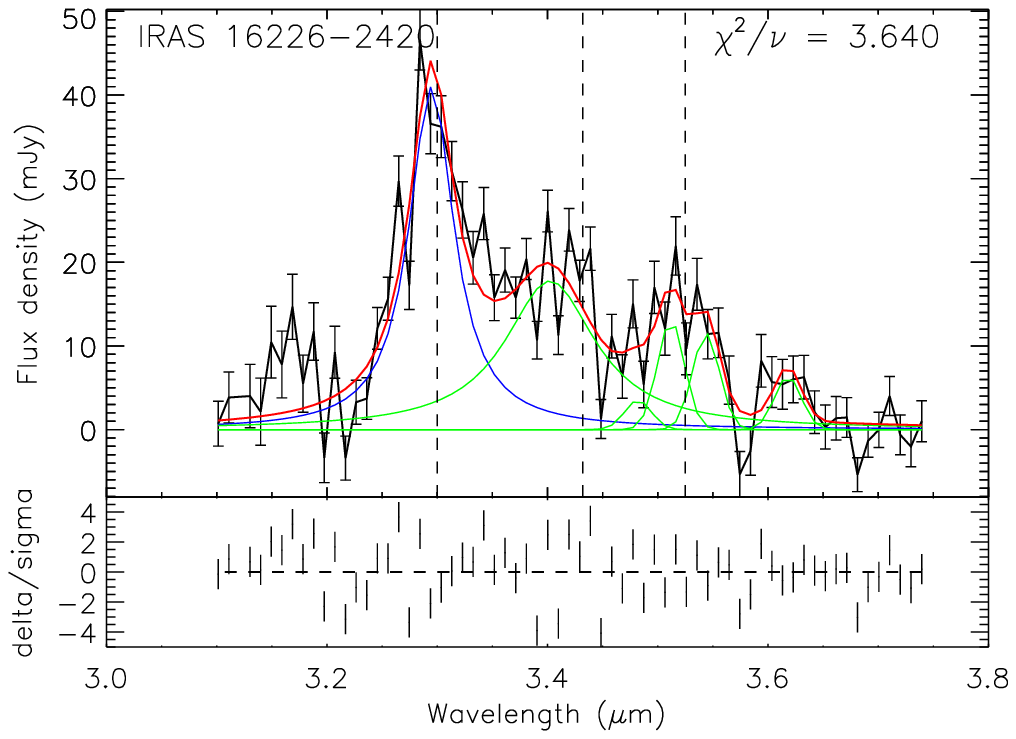}
		\includegraphics[width=0.8\linewidth]{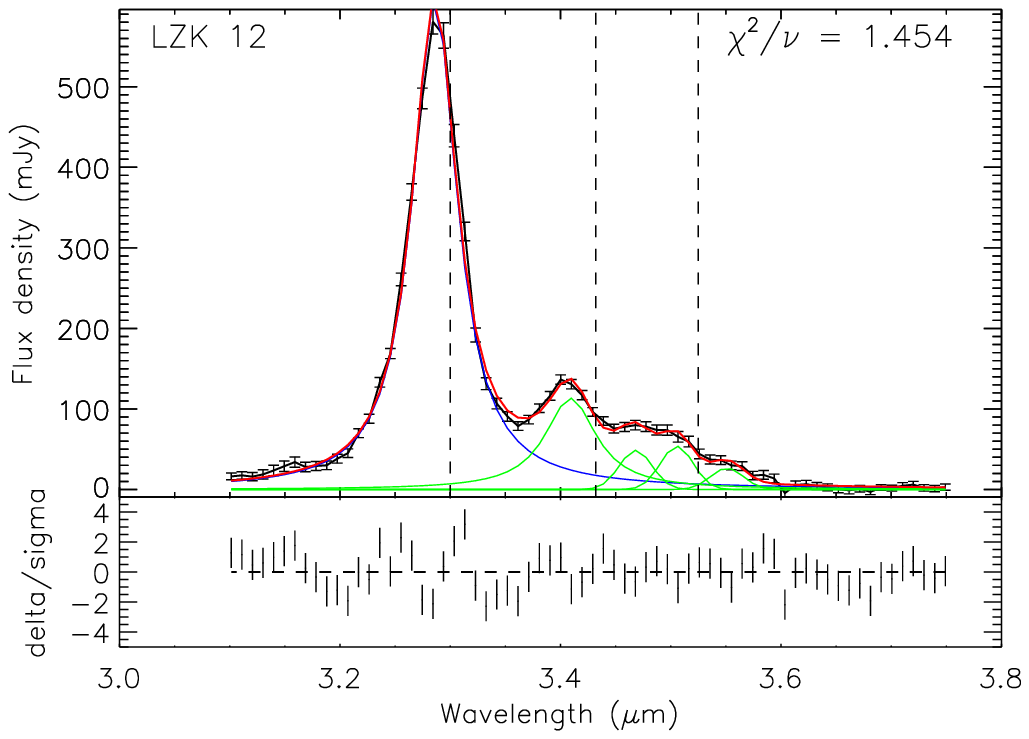}
	\end{minipage}
	\begin{minipage}{.45\linewidth}
		\centering
		\includegraphics[width=0.8\linewidth]{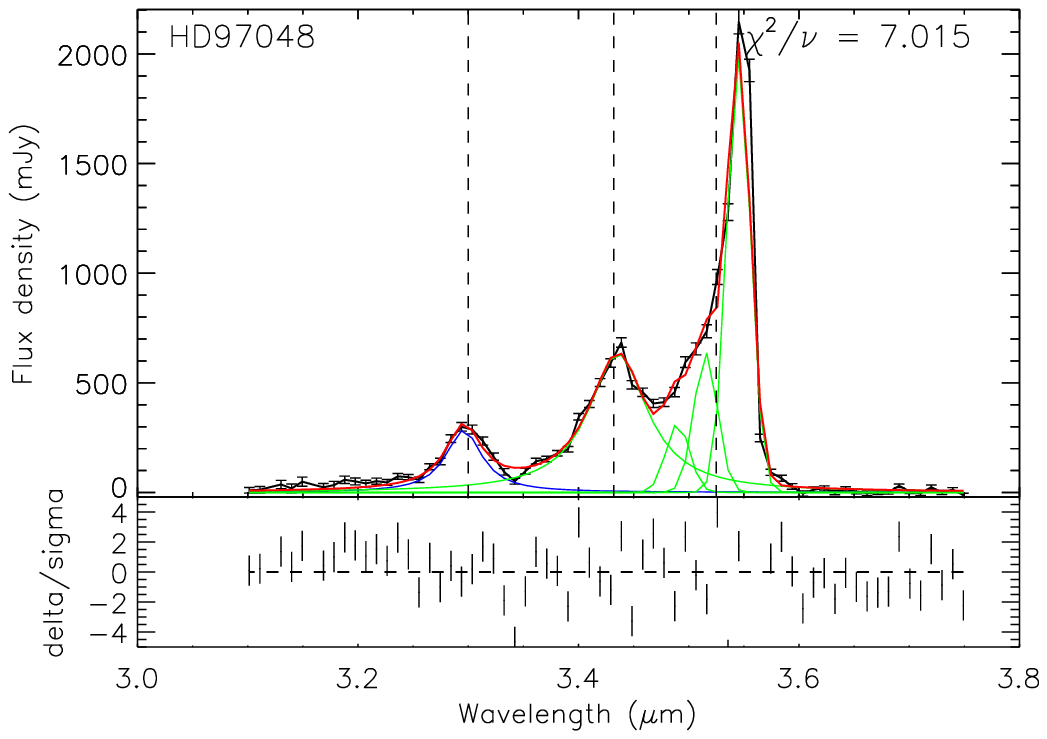}
		\includegraphics[width=0.8\linewidth]{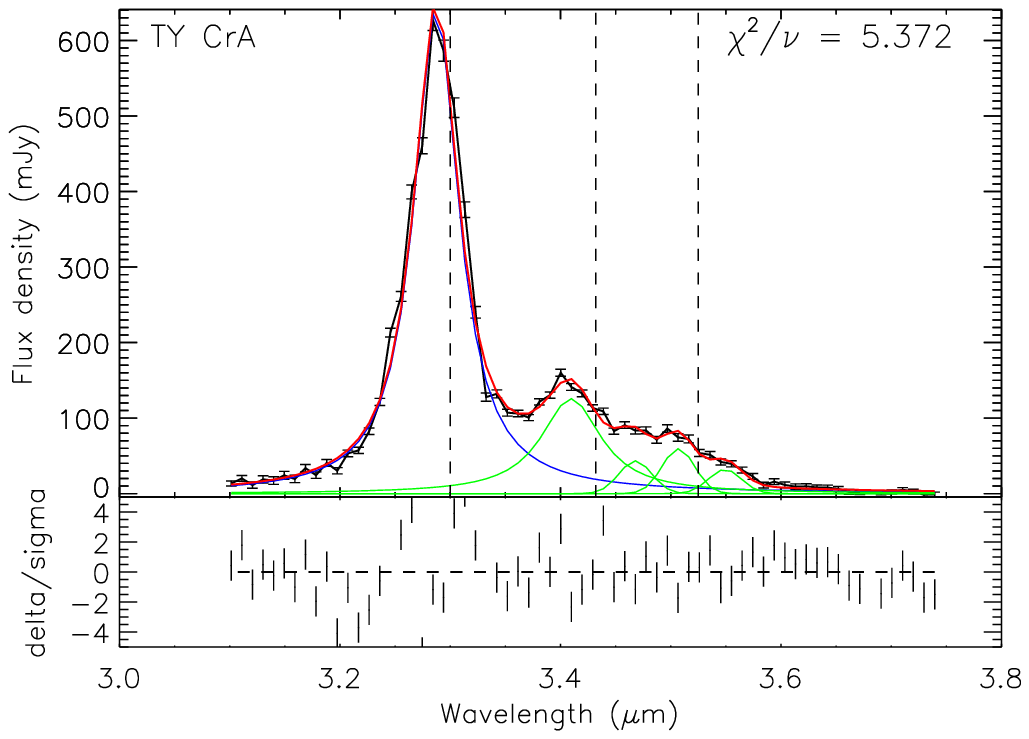}
		\includegraphics[width=0.8\linewidth]{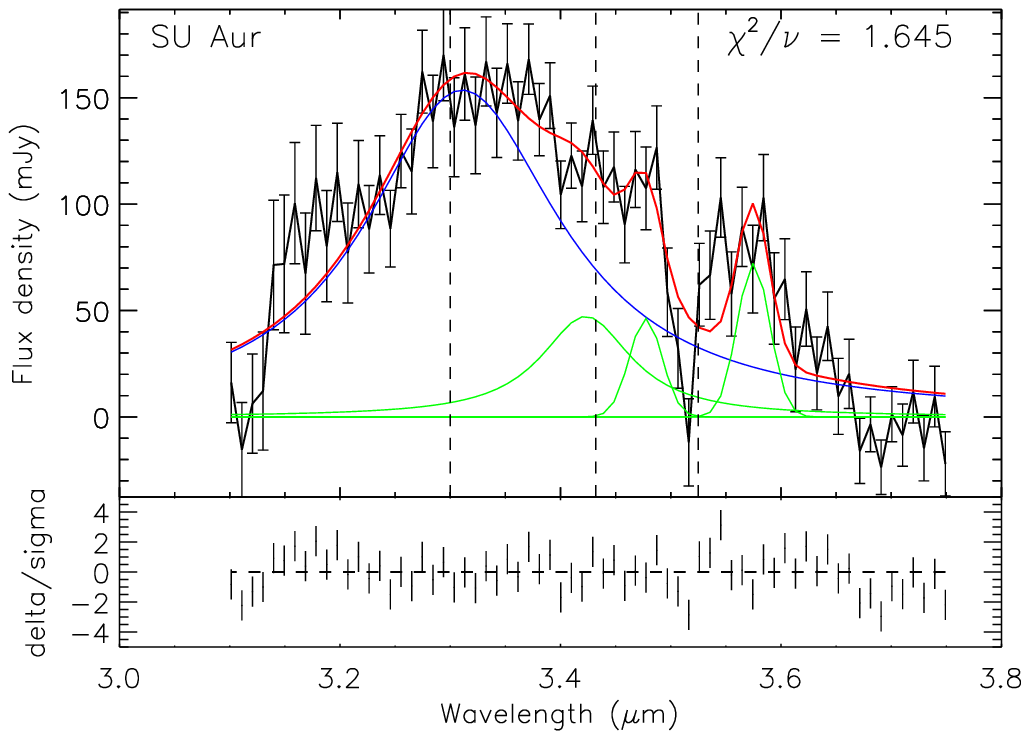}
		\includegraphics[width=0.8\linewidth]{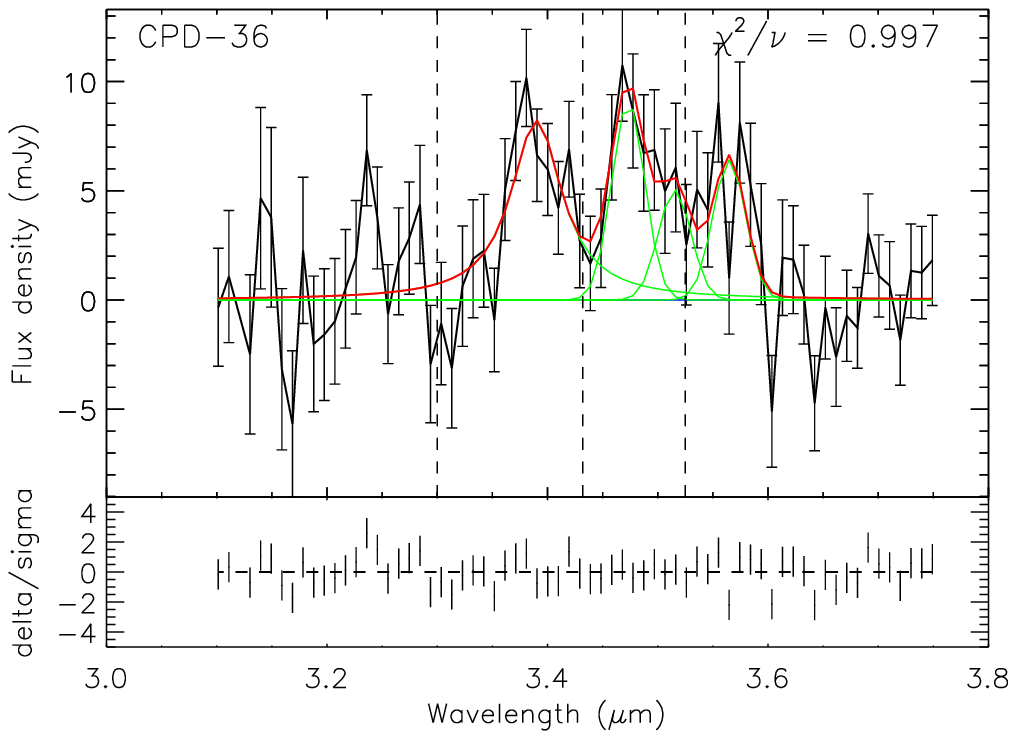}
		\includegraphics[width=0.8\linewidth]{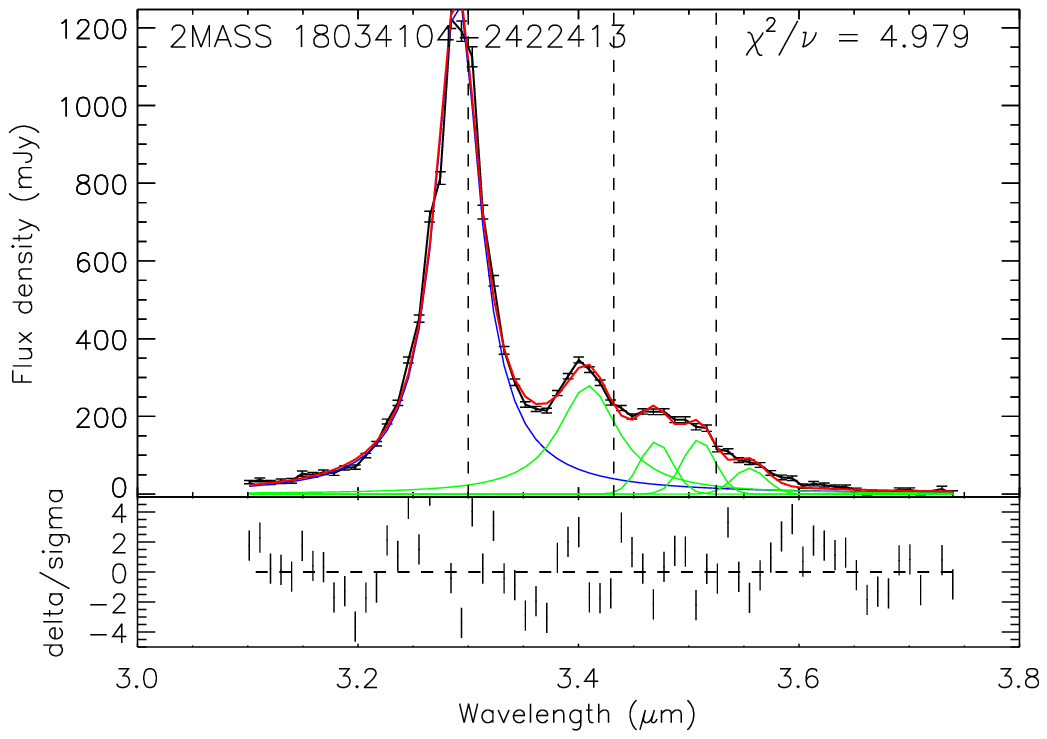}
	\end{minipage}
	\vspace{3mm}
	
	\caption{Results of spectral fitting to the hydrocarbon dust features for the central 10\farcs5 region of each YSO. The line colors and the styles are the same as in figure 1. The dashed lines correspond to the peak wavelengths of the main hydrocarbon dust features seen in the spectrum of Elias 1. The fitting residuals normalized by the errors are shown in the bottom panels. {Alt text: Ten panels showing the results of the spectral fitting.}} 
	\label{figure3}	
\end{figure*}

Figure \ref{figure2}(a) shows the relationship between the aromatic and aliphatic feature fluxes for the inner 10\farcs5 region of each YSO. The solid line ($y=0.3x$) in the figure represents the relation typical of the interstellar hydrocarbon dust in a star-forming galaxy (\cite{kon26}). In comparison with the line, most of the targets show significantly higher aliphatic to aromatic feature flux ratios, with S/N $\ge$ 4 for both vertical and horizontal axes, while TY CrA, LZK 12, and 2MASS J18034104-2422413 exhibit  feature flux ratios similar to the typical value. In particular, Elias 1 and HD97048 display exceptionally high feature flux ratios, which exhibit the strong aliphatic features due to nanodiamonds, as already reported in the previous studies (e.g., \cite{vank02}; \cite{goto09}). 

\begin{table*}[htb]
	\centering
        \caption{Fitting results for the aliphatic emission feature fluxes.}
	\label{tab:ali2341}
		\begin{tabular}{lllll}
		\hline
		Object name &\multicolumn{2}{c}{Inner region} &\multicolumn{2}{c}{Outer region$^*$} \\
			& Aliphatic 1 & Aliphatic 2$+$3$+$4 & Aliphatic 1 & Aliphatic 2$+$3$+$4 \\
			& \multicolumn{4}{c}{($\times 10^{-15}$ W m$^{-2}$)} \\
		\hline\hline
		Elias 1 & 3.04 $\pm$ 0.12 & 1.91 $\pm$ 0.12 & 0.053 $\pm$ 0.003 & 0.096 $\pm$ 0.005 \\
		HD97048 & 16.31 $\pm$ 0.21 & 18.8 $\pm$ 0.27 & 0.625 $\pm$ 0.009 & 0.662 $\pm$ 0.009 \\
		BD+40 4124 & 0.841 $\pm$ 0.026 & 0.206 $\pm$ 0.022 & 0.325 $\pm$ 0.006 & 0.095 $\pm$ 0.005  \\
		TY CrA & 3.377 $\pm$ 0.055 & 1.227 $\pm$ 0.044 & 2.647 $\pm$ 0.014 & 0.736 $\pm$ 0.009 \\
		CD-42 11721 & 16.66 $\pm$ 0.47& 4.30 $\pm$ 0.30 & 5.03 $\pm$ 0.09 & 1.40 $\pm$ 0.06 \\
		SU Aur & 1.90 $\pm$ 0.33 & 1.06 $\pm$ 0.16 & 0.00 & 0.00 \\
		IRAS 16226-2420 & 0.723 $\pm$ 0.038 & 0.491 $\pm$ 0.030 & 0.495 $\pm$ 0.010 & 0.124 $\pm$ 0.006 \\
		CPD-36 6759 & 0.190 $\pm$ 0.024 & 0.186 $\pm$ 0.026 & 0.00 & 0.00 \\
		LZK 12 & 2.594 $\pm$ 0.063 & 1.178 $\pm$ 0.058 & 1.926 $\pm$ 0.015 & 0.722 $\pm$ 0.011 \\
		2MASS J18034104-2422413 & 7.452 $\pm$ 0.088 & 3.106 $\pm$ 0.070 & 3.86 $\pm$ 0.04 & 1.69 $\pm$ 0.04 \\
	\hline
		\end{tabular}
		\begin{tabnote}
\footnotemark[$*$] Fluxes multiplied by a factor of 7/23, considering the difference in the slit aperture size between the inner and outer regions.
\end{tabnote}

\end{table*}

\begin{figure*}
 \centering
 \begin{minipage}[b]{0.49\linewidth}
  \centering
  \includegraphics[width=10cm]{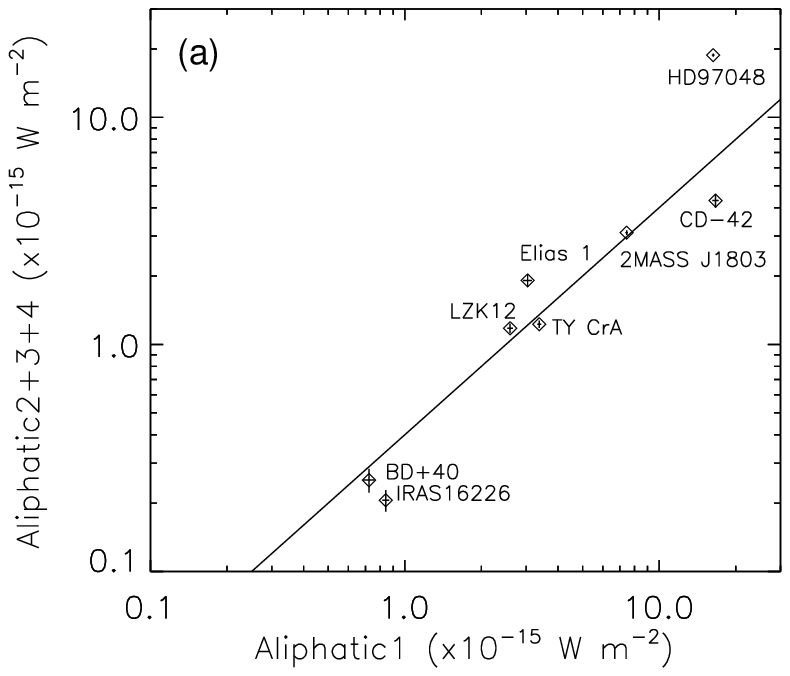} 
\end{minipage}
\begin{minipage}[b]{0.49\linewidth}
 \centering
  \includegraphics[width=10cm]{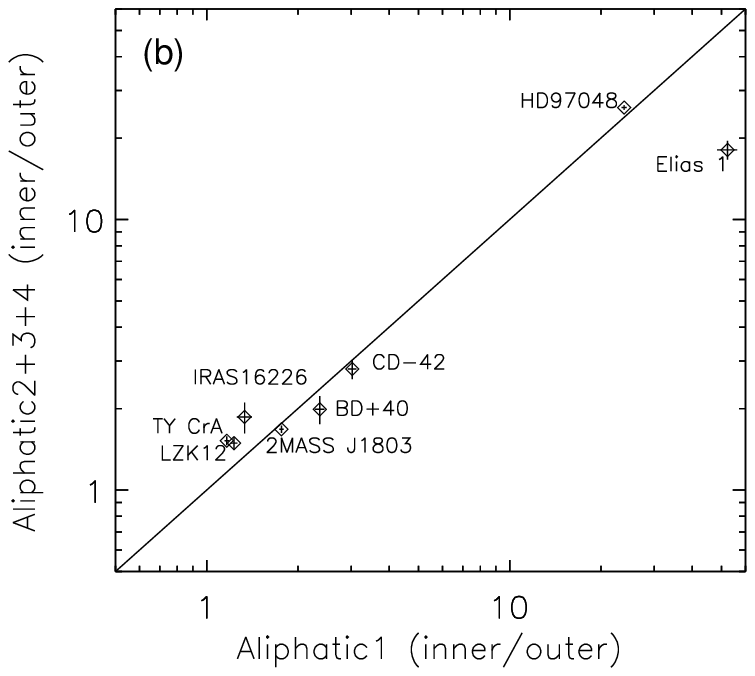} 
\end{minipage}
\caption{Same as figure 2, but plotted for the relationships between the aliphatic 2$+$3$+$4 and aliphatic1 component fluxes. The solid line in panel (a) represents the relation typical of the interstellar hydrocarbon dust in a star-forming galaxy ($y=0.4x$; \cite{kon26}).  The solid line in panel (b) indicates $y=x$, where the shape of the aliphatic feature profile remain unchanged from the outer to the inner regions. {Alt text: Two scatter plots.}}
\label{figure4}
\end{figure*}

Figure \ref{figure2}(b) shows the relationship between the increases in the aromatic and aliphatic feature intensities from the outer (10\farcs5$-$45\farcs0) to the inner ($\le$ 10\farcs5) region. Here, the feature intensities are the fluxes divided by the slit aperture areas for the inner and outer regions, and thus the coordinates of (1, 1) correspond to no increase in their intensities by the central YSO activity. SU Aur is not shown in the plot, for which the aromatic features are not significantly detected in the outer region. As seen in the figure, Elias 1 and HD97048 show increases in the aromatic and aliphatic feature intensities much larger than the other targets due to the heating by the central YSOs, both of which are known to exhibit the nanodiamond features. The solid line indicates $y=x$, where the intensity ratio of the aliphatic to aromatic features remain unchanged from the outer to the inner regions. As seen in the figure, most of the targets lie approximately along this line, indicating that both aromatic and aliphatic hydrocarbons are heated similarly due to the YSO activity. More precisely, however, HD97048, CD-42 11721, IRAS 16226-2420, and 2MASS J18034104-2422413 show significantly higher flux ratios of the aliphatic to aromatic features in the inner region (S/N $\ge$ 4 for both vertical and horizontal axes), while TY CrA shows a lower flux ratio. Therefore, the hydrocarbon dust in these 5 targets at least is likely to have changed its spectral features and thus have undergone processing by the YSO activity.

Table \ref{tab:lineice} summarizes the fitting results for the fluxes of the hydrogen recombination Br$\alpha$ line and the equivalent widths (EWs) of the H$_2$O and the CO$_2$ ice absorption features in the inner region of each target. For the Br$\alpha$ line emission, the continuum fitting in the corresponding wavelength range was not satisfactory for 3 targets (BD+40 4124, CD-42 11721, and IRAS 16226-2420); therefore, the line fluxes were re-derived using a local wavelength range (3.95$-$4.15 \textmu m) with a linear plus Gaussian function fitting for those targets. The Br$\alpha$ emission is significantly detected (S/N $\ge$ 5) in 5 targets among the sample: Elias 1, HD97048, BD+40 4124, TY CrA, and 2MASS J18034104-2422413. In particular, 2MASS J18034104-2422413 shows very strong Br$\alpha$ emission, and the presence of multiple hydrogen recombination lines is clearly identified in its spectrum. As for the H$_2$O and the CO$_2$ ice absorption features, both are significantly and reliably detected from the following 6 targets: Elias 1, HD97048, TY CrA, SU Aur, LZK 12, and 2MASS J18034104-2422413. The other results, however, may be affected by systematic uncertainties in the continuum fitting, potentially leading to spurious feature structures, % the results for the H$_2$O ice in 3 targets (CD-42 11721, IRAS 16226-2420, and CPD-36 6759), and for the CO$_2$ ice in 3 targets (BD+40 4124, CD-42 11721, and IRAS 16226-2420), 
which are excluded from the following discussion. 

Finally, table \ref{tab:lineice} also presents the continuum fluxes at wavelengths of 2.7 \textmu m and 3.9 \textmu m in the inner regions.  We obtain these fluxes by integrating the continuum component of the spectral fitting results over the wavelength ranges of 2.6$-$2.8 \textmu m and 3.75$-$3.95 \textmu m. The former, the 2.7 \textmu m continuum, is considered to be dominated by emission from the central YSO,  while the latter, the 3.9 \textmu m continuum, is likely to contain some contribution from the circumstellar dust emission as seen in figure \ref{figure1}. Hence the flux ratio of the 3.9 \textmu m to 2.7 \textmu m continuum is likely to reflect the relative abundance of the circumstellar dust associated with the central YSO, which will be used in the discussion.

\begin{figure*}[tbp]
	\centering
	\begin{minipage}{0.49\linewidth}
		\centering
		\includegraphics[width=10cm]{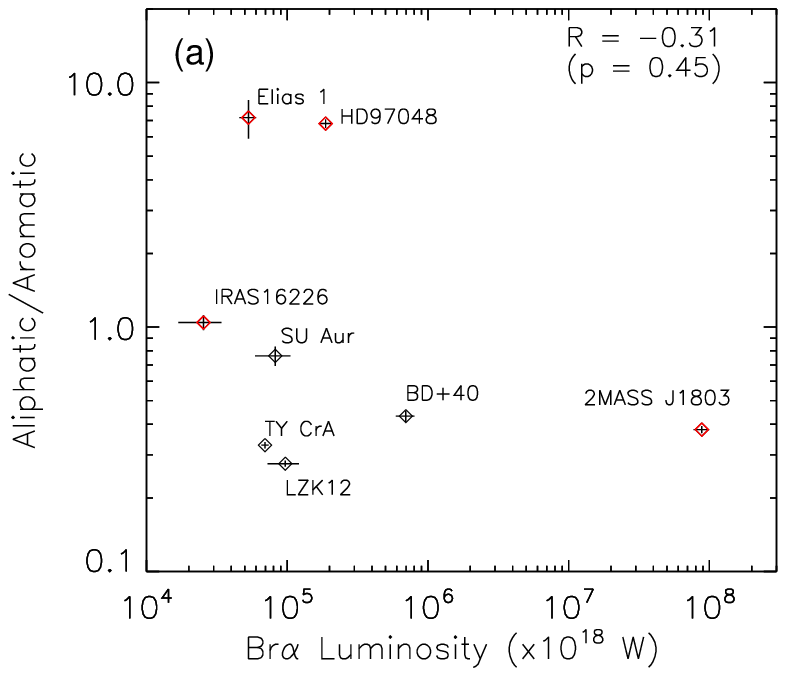}
		\includegraphics[width=10cm]{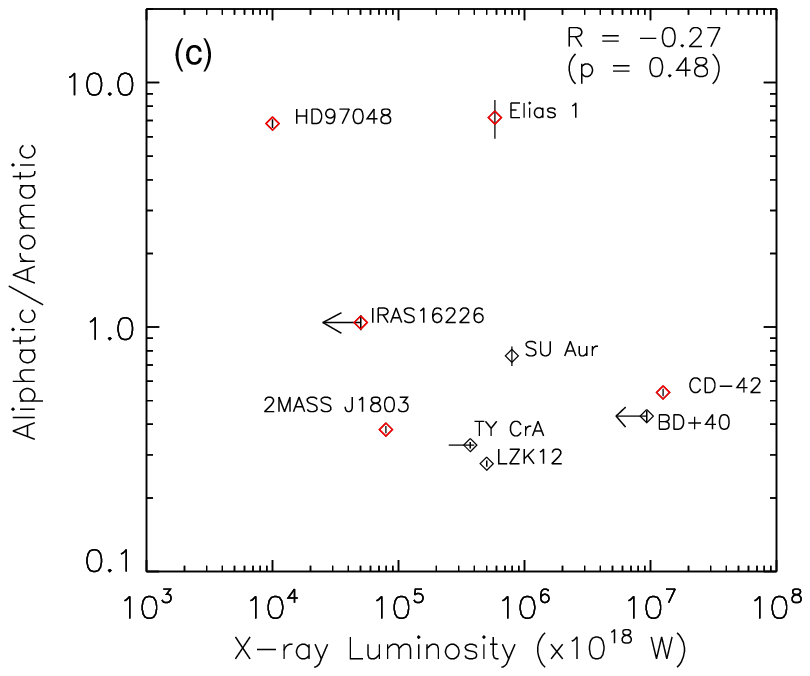}
	\end{minipage}
	\begin{minipage}{.49\linewidth}
		\centering
		\includegraphics[width=10cm]{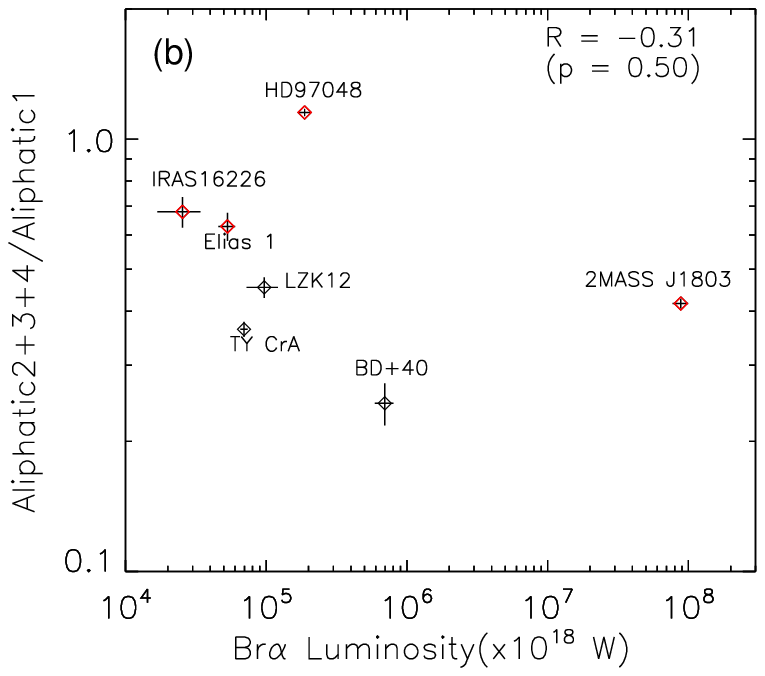}
		\includegraphics[width=10cm]{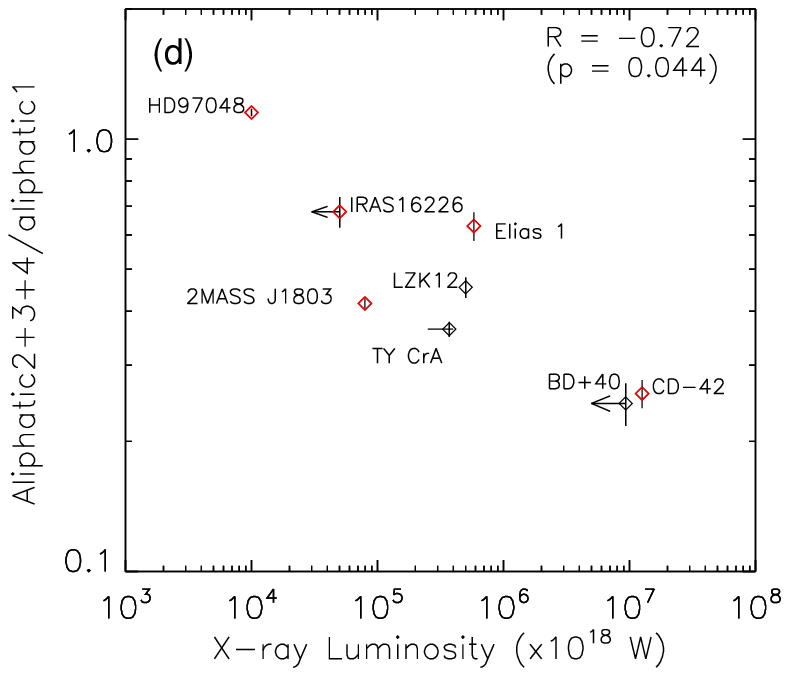}
	\end{minipage}
	
	\vspace{3mm}
	
	\caption{Ratios of (a) the aliphatic to aromatic feature fluxes and (b) aliphatic 2$+$3$+$4 to aliphatic 1 component fluxes plotted against the Br$\alpha$ luminosities. The ratios of (c) the aliphatic to aromatic feature fluxes and (d) the aliphatic 2$+$3$+$4 and aliphatic 1 component fluxes plotted against the X-ray luminosities. Here and hereafter, the correlation coefficients (R) and p-values are shown in the upper right corner. The sources included in the correlation analysis are (a) Elias 1, HD97048, BD+40 4124, TY CrA, SU Aur, IRAS 16226-2420, LZK 12, 2MASS J18034104-2422413, (b) Elias 1, HD97048, BD+40 4124, TY CrA, IRAS 16226-2420, LZK 12, 2MASS J18034104-2422413, (c) Elias 1, HD97048, TY CrA, CD-42 11721, SU Aur, LZK 12, 2MASS J18034104-2422413, and (d) Elias 1, HD97048, TY CrA, CD-42 11721, LZK 12, 2MASS J18034104-2422413. Upper limits are excluded from the analysis. The data points of the targets where the aliphatic to aromatic ratio and/or the aliphatic component ratio change significantly from the outer to the inner region, excluding TY CrA and LZK 12, are shown in red marks. {Alt text: Four scatter plots.}} 
	\label{figure5}	
\end{figure*}

\subsection{Detailed model fitting to the spectral feature profiles of hydrocarbon dust}
\label{ssec:32}

In the previous subsection, model fitting is performed to the full spectra using a complex continuum model that includes the ice absorption features. In this subsection, to examine the aromatic and the aliphatic features in more detail, we extract spectra focusing only on the hydrocarbon dust profiles, as shown in figure \ref{figure3}.  Here the observed spectra of the aromatic plus aliphatic features, $F'_{\nu}$, are derived by the following equation:	
	\begin{equation}
	F'_{\nu}  = F_{\nu} \times \exp(\tau_{\nu}) -  F_{\nu, {\rm gas}} - F_{\nu, {\rm cont}}, 
	\end{equation}
where $\tau_\nu$, $F_{\nu, {\rm gas}}$, and $F_{\nu, {\rm cont}}$ are determined by the overall model fitting to the total spectra in the previous subsection, while $F_\nu$ is the observed flux densities. The vertical dashed lines in the figure approximately correspond to the peak wavelengths of 3 main features observed in the spectrum of Elias 1, which are located at 3.30 \textmu m, 3.43 \textmu m, and 3.53 \textmu m. For Elias 1 and HD97048, it is confirmed that the feature fluxes at wavelengths longer than 3.5 \textmu m are significantly enhanced relative to the shorter-wavelength components, which is consistent with the previous studies indicating strong C-H emission originating from nanodiamonds (\cite{vank02}; \cite{goto09}). The peak of the longest wavelength feature in HD97048 shows a significant offset from that in Elias 1. All the other targets also exhibit significant feature fluxes at wavelengths longer than 3.5 \textmu m. In particular, CD-42 11721 and IRAS 16226-2420 may also show a hint of the unusual aliphatic feature attributable to the nanodiamond emission.

Following the fitting procedure described in section 2, we refit the data using 1 aromatic component (Drude profile) and 4 aliphatic components (1 Lorentzian and 3 Gaussian profiles). Among the 4 aliphatic components, the Lorentzian component on the shorter-wavelength side is defined as aliphatic 1, while the 3 Gaussian components on the longer-wavelength side are grouped as aliphatic 2$+$3$+$4. Table \ref{tab:ali2341} summarizes the fitting results for the aliphatic emission feature fluxes. These values are used in the following section as parameters characterizing the shape of the aliphatic features. For SU Aur, there is a large uncertainty in the separation of the aliphatic 1 from the aliphatic 2$+$3$+$4 components, judging from its spectral shape of the aliphatic feature. Therefore, the results of SU Aur are not included in the following scatter plots related to  the aliphatic component ratio. In addition to the variation in the shape of the aliphatic feature profile, we also find that the central wavelength of the aromatic feature varies between 3.28 \textmu m and 3.30 \textmu m significantly from target to target, as can be recognized by the vertical dashed lines at 3.30 \textmu m in figure \ref{figure3}, the result of which will be discussed later in section \ref{ssec:42}.

	\begin{table*}[htb]
	\centering
        \caption{Summary of the X-ray observations of the present targets}
	\label{tab:x}
		\begin{tabular}{llll}
		\hline
		Object name & log Lx (erg/s) & Satellite & Reference\\
		\hline
		Elias 1 & 31.0 or 31.3 & ASCA & Hamaguchi et al. (2005)\\
		        & 29.96 or 30.8 & Chandra 0.5$-$8 keV & Steizer et al. (2006)\\
		HD97048 & $<$29.8 & ASCA & Hamaguchi et al. (2005)\\
		       & 29 & ROSAT 0.1$-$2.4 keV & Hamaguchi et al. (2005)\\
		BD+40 4124 & $<$31.97 & Einstein 0.16$-$3.5 keV & Damiani et al. (1994)\\
		TY CrA & 30.63 & Chandra 0.5$-$8 keV & Steizer et al. (2006)\\
		         & 30.4$-$30.6 & ASCA & Hamaguchi et al. (2005)\\
		CD-42 11721 &  32.1 & ASCA & Hamaguchi et al. (2005)\\
		SU Aur & 30.9 & XMM 0.3$-$10 keV & Telleschi et al. (2007) \\
		IRAS 16226-2420 & $<$29.7 & ROSAT 0.1$-$2.4 keV & Boller et al. (2016)\\
		CPD-36 6759 & $<$29.3 & ROSAT 0.1$-$2.4 keV & Boller et al. (2016) \\
		LZK 12 & 30.7 & ROSAT & Preibisch et al. (1997)\\
		2MASS J18034104-2422413 & 29.9 & Chandra 0.5$-$8 keV & Broos et al. (2013)\\
	\hline
		\end{tabular}
		
\end{table*}

Figure \ref{figure4}(a) shows the relationship between the aliphatic 2$+$3$+$4 and aliphatic 1 component fluxes for the inner 10\farcs5 region of each YSO. The solid line ($y=0.4x$) represents the relation typical of the interstellar hydrocarbon dust in a star-forming galaxy (\cite{kon26}). In comparison with the line, most of the targets follow the typical relation, except that Elias 1 and HD97048, which are known to exhibit the nanodiamond features, show significantly higher aliphatic 2$+$3$+$4 fluxes, while CD-42 11721 and IRAS 16226-2420 lower aliphatic 2$+$3$+$4 fluxes, with S/N $\ge$ 4 for both vertical and horizontal axes. Figure \ref{figure4}(b) shows the relationship between the increases in the aliphatic 2$+$3$+$4 and aliphatic 1 component intensities from the outer to the inner region. Similarly to figure \ref{figure2}(b), Elias 1 and HD97048 show increases much larger than the other targets. The solid line, again, indicates $y=x$, where the shape of the aliphatic feature profile remains unchanged from the outer to the inner regions. In comparison with the relation expected for no variation in the shape of the aliphatic feature profile, most of the targets lie approximately along this line as seen in the figure. More precisely, however, TY CrA and LZK12 show significantly higher flux ratios of the aliphatic 2$+$3$+$4 to aliphatic 1 components in the inner regions (S/N $\ge$ 4 for both vertical and horizontal axes), while Elias 1 shows a lower flux ratio. Therefore, the hydrocarbon dust in these 3 targets at least is likely to have changed its spectral features and thus have undergone processing by the YSO activity.

\section{Discussion}\label{sec:4}
%%議論

\subsection{Relationship with the properties of the central YSOs }\label{ssec41}

We find that the spectral feature profiles of the C-H emission from hydrocarbon dust in the circumstellar environments of YSOs vary significantly among the 9 targets which are significantly detected in the PAH emission. In particular, Elias 1 and HD97048 exhibit prominent C-H emission from hydrogenated nanodiamonds, while other targets such as CD-42 11721 and IRAS 16226-2420 also show distinctive spectral features. To characterize these variations in the feature profiles, we focus on  the following 2 parameters: the flux ratio of aliphatic to aromatic feature (hereafter, aliphatic to aromatic ratio) and the flux ratio of aliphatic 2$+$3$+$4 to aliphatic 1 feature (aliphatic component ratio) as defined and obtained in section \ref{sec:3}. Compared with the values typical of the interstellar hydrocarbon dust in a star-forming galaxy, the aliphatic to aromatic ratio is significantly higher for 7 out of the 9 targets, while the other 2 targets show similar values. On the other hand, the aliphatic component ratio is similar to the typical value for 5 out of the 9 targets, while another 2 targets show significantly higher values but the other 2 targets lower values. We investigate how these parameters relate to the properties of the central YSOs. As shown in figures \ref{figure2}(b) and \ref{figure4}(b), either or both of these parameters change significantly from the outer to the inner regions for Elias 1, HD97048, CD-42 11721, TY CrA, IRAS 16226-2420, LZK 12, and 2MASS J18034104-2422413, and therefore the properties of the hydrocarbon dust are likely to be affected by the activity of, at least, these 7 YSOs. 

It should be noted, however, that previous studies with Spitzer have shown that the PAH emission is frequently associated with reflection nebulae or the broader circumstellar medium for B-type YSOs (e.g., \cite{aru23}). In this regard, TY CrA and LZK 12 in our sample, which are B-type YSOs (see table~1), show aliphatic to aromatic ratios and aliphatic component ratios both typical of the interstellar hydrocarbon dust (figures~2(a), 4(a)), and also LZK 12 does not show a significant change in the aliphatic to aromatic ratio from the outer to the inner region (figure~2(b)). Hence, among the 7 YSOs, we consider that a significant fraction of the hydrocarbon dust emission from TY CrA and LZK 12 may originate in extended nebulosity. The remaining 5 targets are more likely to reveal the properties of the circumstellar hydrocarbon dust associated with the central YSOs, although it is difficult to distinguish whether the emission originates from the circumstellar disk or the envelope, given the spatial resolution of AKARI. In the following plots (figures 5$-$8), the data points of these 5 targets more likely to be affected by the YSO activity are highlighted in red marks.

\begin{figure*}[tbp]
	\centering
	\begin{minipage}{0.49\linewidth}
		\centering
		\includegraphics[width=10cm]{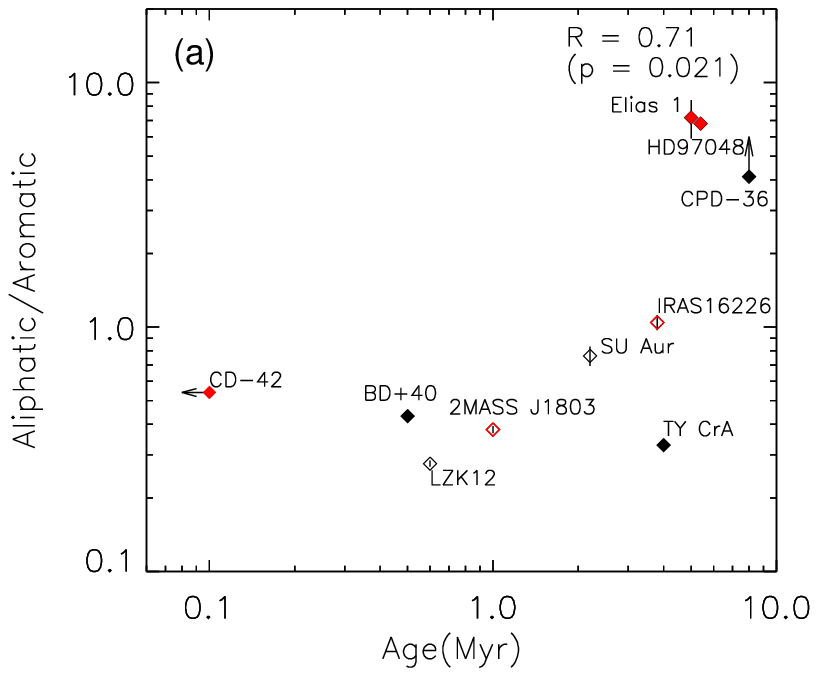}
		\includegraphics[width=10cm]{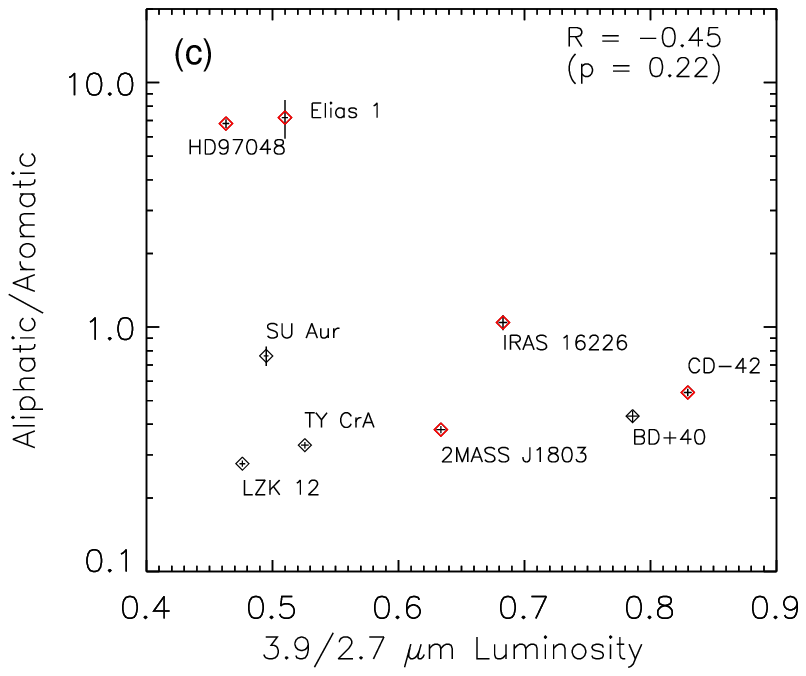}
	\end{minipage}
	\begin{minipage}{.49\linewidth}
		\centering
		\includegraphics[width=10cm]{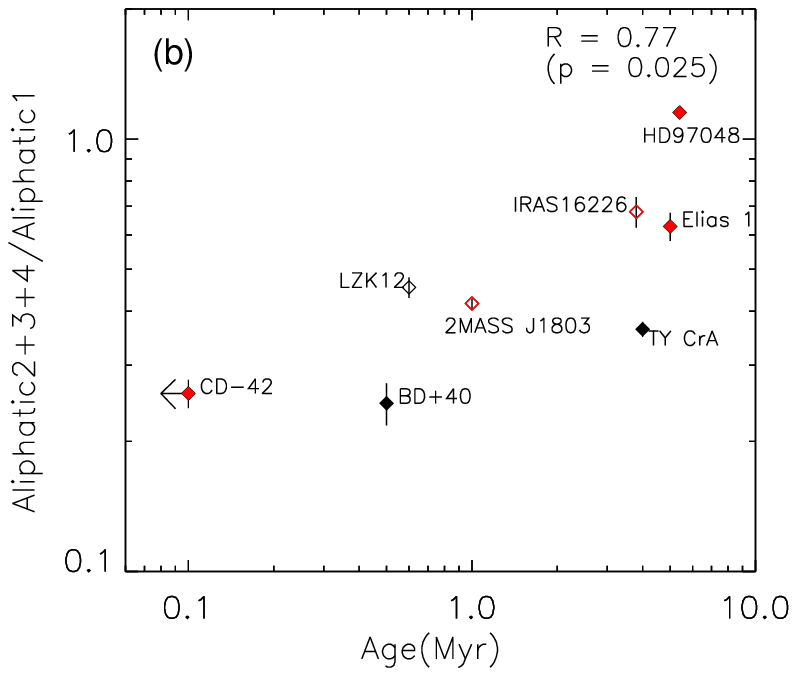}
		\includegraphics[width=10cm]{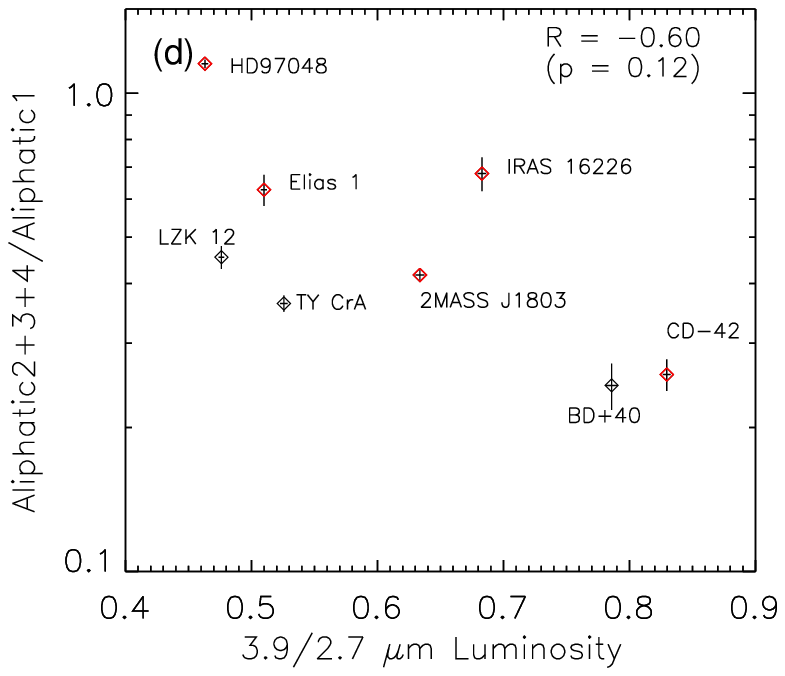}
	\end{minipage}
	
	\vspace{3mm}
	
	\caption{Ratios of (a) the aliphatic to aromatic feature fluxes and (b) the aliphatic 2$+$3$+$4 to aliphatic 1 component fluxes plotted against the YSO age. The ratios of (c) the aliphatic to aromatic feature fluxes and (d) the aliphatic 2$+$3$+$4 to aliphatic 1 component fluxes plotted against the 3.9 to 2.7 $\mu$m continuum flux ratio. The sources included in the correlation analysis are (a) Elias 1, HD97048, BD+40 4124, TY CrA, SU Aur, IRAS16226-2420, LZK 12, 2MASS J18034104-2422413, (b) Elias 1, HD97048, BD+40 4124, TY CrA, IRAS16226-2420, LZK 12, 2MASS J18034104-2422413, (c) Elias 1, HD97048, BD+40 4124, TY CrA, CD-42 11721, SU Aur, IRAS16226-2420, LZK 12, 2MASS J18034104-2422413, and (d) Elias 1, HD97048,  BD+40 4124, TY CrA, CD-42 11721, IRAS16226-2420, LZK 12, 2MASS J18034104-2422413. Upper limits are excluded from the analysis. The data points of the targets where the aliphatic to aromatic ratio and/or the aliphatic component ratio change significantly from the outer to the inner region, excluding TY CrA and LZK 12, are shown in red marks, while those of the Herbig Ae/Be stars are indicated by filled marks in panels (a) and (b). {Alt text: Four scatter plots.}} 
	\label{figure6}	
\end{figure*}

Figure \ref{figure5} shows the relationships of these flux ratios with the Br$\alpha$ and X-ray luminosities, while figure \ref{figure6} presents their relations with the YSO age listed in table \ref{tab:target} and the intensity ratio of the 2.7 \textmu m to the 3.9 \textmu m continuum obtained in section \ref{ssec:31}. Using the distances to the targets listed in table~1, the measured fluxes are converted into the luminosities. In the following discussion, we refer to the correlation coefficients and their p-values shown in each scatter plot, which are evaluated using Pearson's correlation test, excluding cases where an apparent correlation is driven only by one to two data points. Upper limits are shown for reference only and not included in the statistical calculations. We adopt thresholds on $R$ ($|R|>0.6$) and p-values ($<0.05$) to judge whether the correlation is significant or not, although the number of the data points is not sufficient in our sample. As seen in figures \ref{figure5}(a) and (b), no correlation is found in the scatter plots related to the Br$\alpha$ luminosity which is expected to reflect the overall general YSO activity. For example, Elias 1 and HD97048, where the nanodiamond emission is detected, do not show relatively high Br$\alpha$ luminosities. Hence the results in figures \ref{figure5} (a) and (b) indicate that the general YSO activity producing ionizing photons is not likely to be a main driver of the processing of the circumstellar hydrocarbon dust.

Table \ref{tab:x} summarizes the X-ray luminosities compiled from the literature for each target, from which we create the plots in figures \ref{figure5}(c) and (d), where the median, the maximum, and the minimum correspond to the X-ray luminosity, and the ends of the error bar, respectively, for the targets with multiple X-ray observational results. Considering the relationship between these X-ray luminosities and the 2 parameters characterizing the variations of the hydrocarbon dust feature profiles, even the targets that suggest strong processing with high-energy irradiation, Elias 1 and HD97048, again, do not show relatively high X-ray luminosities. Although figure \ref{figure5}(d) suggests correlation between the aliphatic component ratio and  the X-ray luminosity, YSOs of higher X-ray luminosities (e.g., CD-42 11721) do not show large deviations from normal aliphatic component ratios ($\sim$0.4) typical of the interstellar hydrocarbon dust (\cite{kon26}). Therefore, no clear evidence is obtained for a direct correlation between the overall high-energy activity of the YSOs related to the X-ray luminosity and the processing of hydrocarbon dust.

On the other hand, as seen in figures \ref{figure6}(a) and (b), both the aliphatic to aromatic ratio and the aliphatic component ratio show significant correlations with the YSO age, where both flux ratios tend to increase with the YSO age. Here, the YSO age of CD-42 11721 is expressed as  the upper limit of 0.1 Myr in the plot, since the estimated age is too young, as shown in table 1. Furthermore, in the relatively old target CPD-36 6759, the flux of the hydrocarbon dust feature itself is very low and the aromatic component is not significantly detected, yielding only a lower limit to the aliphatic to aromatic ratio. Since the YSO age is coupled with the YSO mass, we indicate the YSOs categorized as Herbig Ae/Be stars (see table~1) with filled marks in the plots, from which we confirm that the observed trend remains even when only the Herbig Ae/Be stars are considered, although the statistics are small. These suggest that there may exist a characteristic evolutionary stage of YSOs that is favorable for the processing of hydrocarbon dust. As seen in figure \ref{figure6}(d), the aliphatic component ratio also shows a decreasing trend with the flux ratio of the 3.9 \textmu m to 2.7 \textmu m continuum, although they are not significantly correlated (i.e., p-value $>0.05$). This continuum ratio is likely to reflect relative contributions of the emission from circumstellar dust, as mentioned in section 3.1. The observed decreasing trend therefore suggests that the aliphatic hydrocarbons tend to undergo stronger processing when the remaining amount of the circumstellar dust is smaller. Hence the YSO age and the amount of the dust associated with the YSO consistently suggest that the later evolutionary stages are favorable for the processing of the hydrocarbon dust.
%Hence the YSO age and the amount of the dust associated with the YSO consistently point toward the later evolutionary stages which are favorable for the processing of the hydrocarbon dust. 

\subsection{Relationship with the properties of the circumstellar ices and aromatic hydrocarbons}\label{ssec:42}
%%空間変化

\begin{figure*}
 \centering
 \begin{minipage}{0.49\linewidth}
  \centering
  \includegraphics[width=10cm]{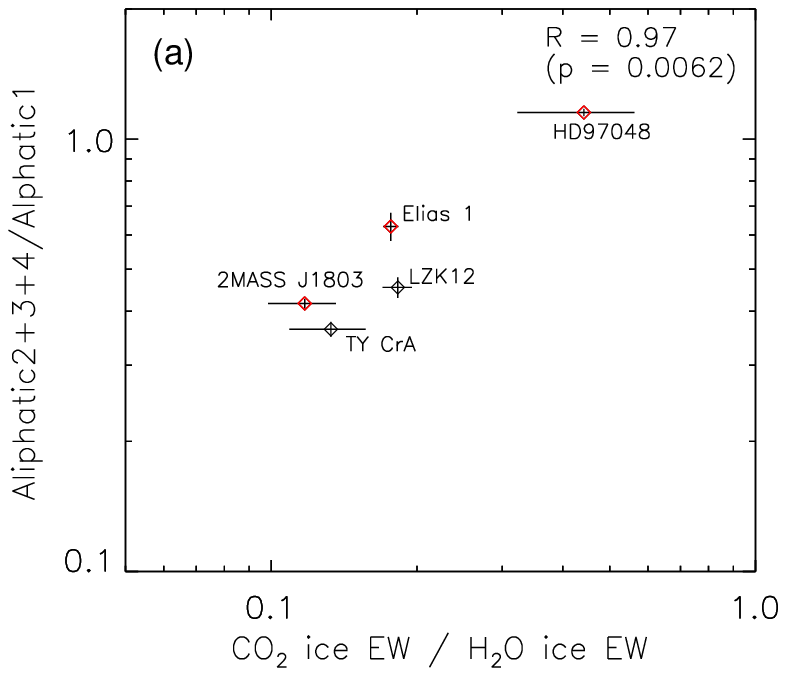} 
\end{minipage}
\begin{minipage}{0.49\linewidth}
 \centering
  \includegraphics[width=10cm]{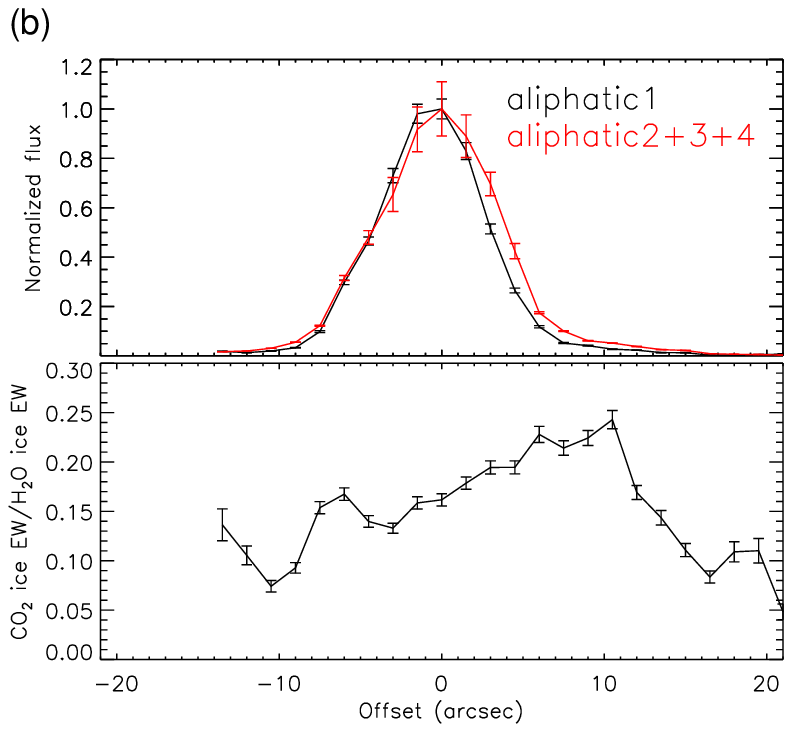}
\end{minipage}
\vspace{3mm}
    \caption{(a) Ratio of the aliphatic 2$+$3$+$4 to the aliphatic 1 component flux plotted against the ratio of the CO$_2$ ice to H$_2$O ice equivalent width (EW). The sources included in the correlation analysis are Elias 1, HD97048, TY CrA, LZK 12, and 2MASS J18034104-2422413. The data points of the targets where the aliphatic to aromatic ratio and/or the aliphatic component ratio change significantly from the outer to the inner region, excluding TY CrA and LZK 12, are shown in red marks. (b) The normalized fluxes of the aliphatic 1 and the aliphatic 2$+$3$+$4 component of Elias 1 are shown  in the upper panel and the ratio of the CO$_2$ ice to the H$_2$O ice EW in the lower panel, both plotted as a function of the positional offset from the center which is defined as the peak position of the continuum intensity. {Alt text: One scatter plot and two projection plots.}}
\label{figure7}
\end{figure*}

Figure \ref{figure7}(a) shows the aliphatic component ratios as a function of the EW ratio of the CO$_2$ to the H$_2$O ice absorption feature for the 5 targets, which suggests that both parameters may be related to each other, although the statistics are rather poor. Among the targets, we focus on Elias 1, which exhibits the extended nanodiamond-like emission, to examine its detailed spatial variation. The other targets have insufficient spatial resolution for AKARI, making Elias 1 the most suitable target for this analysis. The spatial variations of the aliphatic 1 and aliphatic 2$+$3$+$4 components are shown in the upper of figure \ref{figure7}(b), each normalized to their peak values. The figure shows that the flux of the aliphatic  2$+$3$+$4 component is significantly enhanced on one side, in the offset range of approximately 3" to 10" from the central YSO, where the EW ratio of the CO$_2$ to the H$_2$O ice absorption feature is also elevated over a similar spatial range, as shown in the lower panel of figure \ref{figure7}(b) . Hence figures \ref{figure7}(a) and (b) both consistently suggest that the enhancement in the abundance of the CO$_2$ ice relative to the H$_2$O ice may be related to the degree of the processing of the aliphatic hydrocarbons. One possible interpretation is that high-energy particles produced through the YSO activity penetrate  circumstellar clouds and promote both the formation of the CO$_2$ ice (\cite{yama12}) and the processing of the aliphatic hydrocarbons. These results suggest that high-energy particles, rather than X-rays, may play a crucial role in the formation of nanodiamonds through the processing of hydrocarbon dust which takes place rather locally.

\begin{figure*}
 \centering
 \begin{minipage}[b]{0.49\linewidth}
  \centering
  \includegraphics[width=10cm]{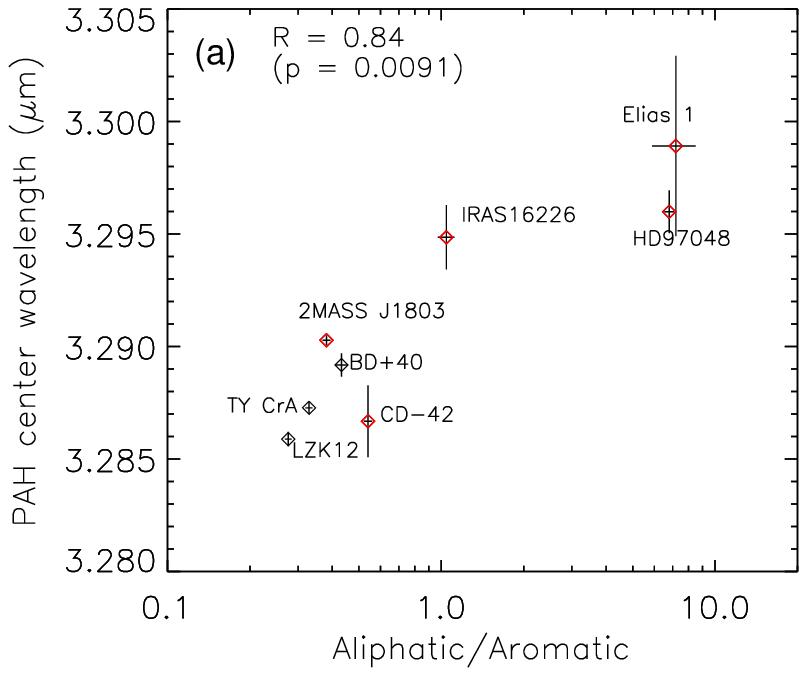} 
\end{minipage}
\begin{minipage}[b]{0.49\linewidth}
 \centering
  \includegraphics[width=10cm]{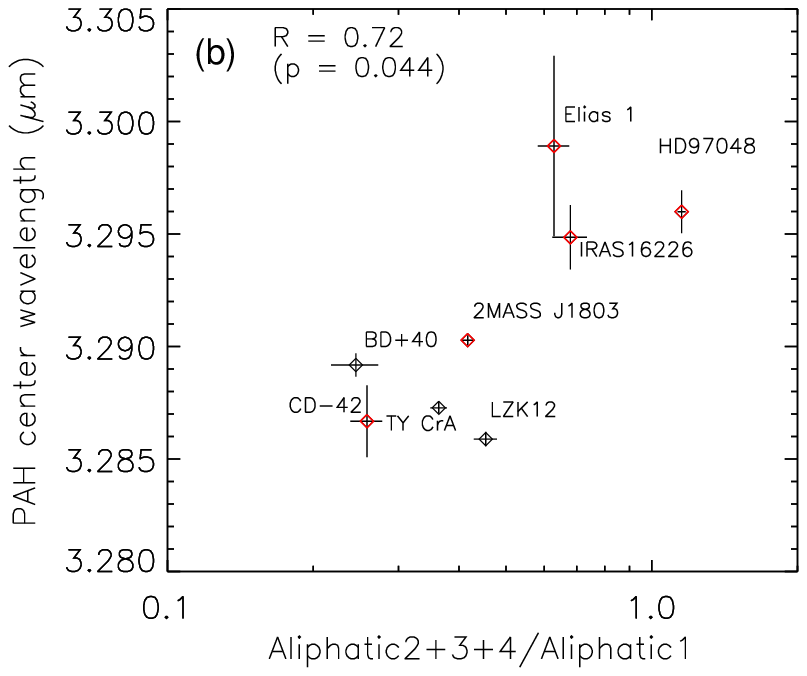} 
\end{minipage}
\caption{PAH central wavelength plotted against (a) the ratio of the aliphatic to the aromatic feature flux and (b) the ratio of the aliphatic 2$+$3$+$4 to the aliphatic 1 component flux. The sources included in the correlation analysis are Elias 1, HD97048, BD+40 4124, TY CrA, CD-42 11721, IRAS16226-2420, LZK 12, 2MASS J18034104-2422413 for both (a) and (b). The data points of the targets where the aliphatic to aromatic ratio and/or the aliphatic component ratio change significantly from the outer to the inner region, excluding TY CrA and LZK 12, are shown in red marks. {Alt text: Two scatter plots.
}}
\label{figure8}
\end{figure*}

We also find that the central wavelength of the aromatic feature varies significantly within the range of 3.28 \textmu m to 3.30 \textmu m, depending on the target, as seen by the dashed lines in figure \ref{figure3}. We therefore examine the relationship between the shifts in the central wavelength and the variations in the hydrocarbon feature profile. Figure \ref{figure8} shows that the central wavelength of the aromatic feature is significantly correlated with both the aliphatic to aromatic ratio and the aliphatic component ratio. The 3.28 \textmu m feature is known to correspond to the C-H vibrational mode in the bay structure of PAHs and is commonly observed in the typical interstellar environments (\cite{can14}). The absence of the significant 3.28 \textmu m feature in some of the targets (e.g., Elias 1, HD97048, and IRAS16226-2420) supports the interpretation that aromatic hydrocarbons in these sources have undergone strong processing, leading to their structural changes from bay-dominant (i.e., edgy) into non-bay (i.e., smooth) structures. Hence, the systematic shift in the central wavelength of the aromatic feature between 3.28 \textmu m and 3.30 \textmu m is observed in correlation with the degree of the processing of the aliphatic hydrocarbons, indicating that the processing mechanisms of the aromatic and the aliphatic hydrocarbons are somewhat related to each other, which may be consistent with a tentative scenario that some circumstellar PAHs could be converted to nanodiamonds through irradiation by high-energy particles.

\section{Conclusion}\label{sec:5}
We have studied the aromatic and the aliphatic hydrocarbon dust spectral features at wavelengths of 3.3$-$3.6 \textmu m for 10 YSOs, using the AKARI near-IR spectroscopic data. In order to reproduce the hydrocarbon dust features, we constructed a spectral model incorporating multiple components with narrowly variable widths and central wavelengths, which enables a unified quantitative characterization of diverse hydrocarbon feature profiles. We find that both aromatic and aliphatic hydrocarbon features are significantly detected in 9 out of the 10 YSOs, their features showing large variations from target to target. Of these, 2 exhibit spectral profiles typical of the interstellar dust, while the other 7 show distinctive profiles characterized by enhanced aliphatic emission relative to aromatic emission. In particular, Elias 1 and HD97048 display markedly strong aliphatic feature fluxes at wavelengths longer than 3.5 \textmu m, consistent with the previous studies which attribute these features to C-H emission from hydrogenated nanodiamonds (\cite{vank02}; \cite{goto09}). In addition, 2 targets show possible evidence for nanodiamond-related feature emission. As for the parameters characterizing the feature shapes, we focus on the aliphatic to aromatic ratio and the aliphatic component ratio. As a result, we find that in 7 targets, including Elias 1 and HD97048, the hydrocarbon feature profiles of hydrocarbon dust vary systematically and significantly from the outer to the inner regions.

To investigate the origin of the target-to-target variations, we examine relationships between the hydrocarbon feature parameters and the hydrogen recombination Br$\alpha$ luminosity, the X-ray luminosity, the YSO age, and the continuum flux ratio at 3.9 \textmu m to 2.7 \textmu m. As a result,  correlation is found for the YSO age and possibly the continuum flux ratio, suggesting that there may exist an evolutionary stage of YSOs that is favorable for hydrocarbon dust processing. In contrast, no clear positive correlation is found for the indicators of the overall YSO activity, such as Br$\alpha$ and X-ray luminosities. In addition, for 6 targets, we detect both H$_2$O and CO$_2$ ice absorption features, the EW ratio of which is found to be likely related to the changes in the aliphatic feature shape. For Elias 1, we also find spatial variations in the aliphatic feature shape within a target, which are correlated with those in the EW ratio of the CO$_2$ to H$_2$O ice. Both suggest a local connection between the ice processing and the formation of nanodiamonds. Finally, a systematic shift in the central wavelength of the aromatic feature between 3.28 \textmu m and 3.30 \textmu m is observed in correlation with the degree of the processing of the aliphatic hydrocarbons, indicating that the processing mechanisms of the aromatic and the aliphatic hydrocarbons are related to each other. These findings may be consistent with a tentative scenario that some of the PAHs may be converted to nanodiamonds due to the irradiation of high-energy particles.

\begin{ack}
We thank the referee for giving us useful comments. This research is based on observations with AKARI, a JAXA project with the participation of ESA. Part of this work is financially supported by JST SPRING, Grant Number JPMJSP2125. The principal author is grateful for the "THERS Make New Standards Program for the Next Generation Researchers."
\end{ack}

%%%%%%%%%%%%%%%%%%%%%%%%%%%%%%%%%%%%%%%

% Sample Data Availability Statements 
% https://academic.oup.com/pages/open-research/research-data#Data%20Availability%20Statements

\appendix %%%%%%%%%%%%%%%%%%%%%%%%%%%%%%%%%%%%%%%%%%%%%%%%%%%%%%%%

% Any journal's BST file (e.g., apj.bst) can be used as PASJ's BST is unavailable.    
% \bibliographystyle{****}
% \bibliography{****}

\end{document}